\documentclass[ reprint,
 aps,
 pra,
 showkeys,
 nofootinbib
]{revtex4-1}
\usepackage{amsfonts}
\usepackage{amssymb}
\usepackage{amsmath}
\usepackage{graphicx}
\usepackage{subfigure}
\usepackage{dcolumn}
\usepackage{bm}
\usepackage{setspace}
\usepackage[hang]{footmisc}
\usepackage[sort&compress]{natbib}
\usepackage{multirow}
\usepackage[sort&compress]{cleveref}
\usepackage{nomencl}
\usepackage[english]{babel}
\usepackage{xcolor, soul}
\usepackage{marginnote}
\usepackage{verbatim}
\sethlcolor{yellow}

\crefname{equation}{\unskip}{\unskip}
\crefname{figure}{\unskip}{\unskip}
\crefname{section}{\unskip}{\unskip}
\crefname{subsection}{\unskip}{\unskip}
\begin{document}
\let\ref\cref

\title{Sphere and cylinder contact mechanics during slip}

\author{J. Wang}
\affiliation{PGI-1, FZ J\"ulich, Germany, EU}
\affiliation{College of Science, Zhongyuan University of Technology, Zhengzhou 450007, China}
\author{ A. Tiwari}
\affiliation{PGI-1, FZ J\"ulich, Germany, EU}
\affiliation{www.MultiscaleConsulting.com}
\author{ I. Sivebaek}
\affiliation{PGI-1, FZ J\"ulich, Germany, EU}
\affiliation{Department of Mechanical Engineering, Technical University of Denmark,
Produktionstorvet, Building 427, Kongens Lyngby 2800, Denmark}
\affiliation{Novo Nordisk Device R \& D, DK-400 Hiller$\phi$d, Denmark}
\author{ B.N.J. Persson}
\affiliation{PGI-1, FZ J\"ulich, Germany, EU}
\affiliation{www.MultiscaleConsulting.com}

%

\begin{abstract}
Using molecular dynamics (MD) we study the dependency of the contact mechanics on the sliding speed when an elastically
soft slab (block) is sliding on a rigid substrate with a ${\rm sin} (q_0 x)$ surface height profile. The atoms on the
block interact with the substrate atoms by Lennard-Jones potentials. We consider contacts with and without adhesion.
We found that the contact area and the friction force are nearly velocity independent for
small velocities ($v < 0.25 \ {\rm m/s}$) in spite of the fact that the shear stress in the contact area is rather non-uniform.
For the case of no adhesion the friction coefficient is very small. For the case of adhesion the friction coefficient is higher,
and is mainly due to energy dissipation at the opening crack tip, where rapid atomic snap-off events occur
during sliding.

Adhesion experiments between a human finger and a clean glass plate were carried out, and for a dry finger no macroscopic adhesion occurred.
We suggest that the observed decrease in the contact area with increasing shear force results from non-adhesive finger-glass contact mechanics,
involving large deformations of a complex layered material.

\end{abstract}

\maketitle
\makenomenclature


{\bf 1 Introduction}

The contact between an elastic ball (or cylinder) and a flat surface is perhaps the simplest possible contact mechanics problem,
and often used in model studies of adhesion and friction. For a stationary contact (no slip) the adhesive interaction
is well described by the Johnson-Kendall-Roberts (JKR) theory\cite{JKR} which has been tested in great detail.
The JKR theory is based on the minimization
of the potential energy which consists of an elastic deformation energy term $U_1$, and a ball-substrate binding energy term
$-\pi r^2 w$, where $r$ is the radius of the circular contact region and $w$ the energy per unit surface area to separate
two flat surfaces made from the same materials as the ball and the substrate. In many applications contact hysteresis occurs where it
during pull-off $w$ may be much larger than the adiabatic value $w_0$, which would prevail during infinitesimal small pull-off speed.
Similarly, during approach $w$ may be much smaller than the adiabatic value $w_0$\cite{AviDoroBo, AviDoroBo1,AviDoroBo2}.

The adhesive contact between an elastic ball and a flat surface, when the ball is exposed to a tangential force
(e.g., a sliding ball), has been studied experimentally in great detail\cite{France,Chaud,PNAS,PRL},
but for this situation there is no good theoretical understanding, in particular
for the dependency of the contact area on the sliding speed (see Ref. \cite{Am}).
Using the elastic continuum approximation, Savkoor and Briggs\cite{Savkoor}
studied the case when the tangential force $F_x$ is constant, and the contact area is displace uniformly (no slip), and found that this
results in a decrease in the area of contact. However, this model cannot explain the dependency of
the contact area on the sliding speed as observed by Vorvolakos and Chaudhury for a
Polydimethylsiloxane (PDMS) ball on a flat substrate\cite{Chaud}. In this case
one expect a uniform frictional stress in the area of contact rather than the highly non-uniform stress
which prevails in the case studied by Savkoor and Briggs. The sphere-flat contact mechanics with a uniform shear stress
was recently studied by Menga, Carbone and Dini\cite{Menga}, and they found that there is no dependency of the contact area
on the magnitude of the shear stress. This is consistent with the experimental data of Vorvolakos and Chaudhury because the
decrease in the contact area they observe occurs at relative high sliding speed, and appear to be related to some effect
not included in model studies so far (see Sec. 3.3).

This paper is organized as follows: in Sec. 2 we discuss the present theoretical understanding,
based on continuum mechanics, of how the contact area between 
a sphere and a flat surface depends on the normal $F_z$ and tangential $F_x$ applied forces. We consider both no-slip condition for a constant
tangential force $F_x$, and slip with a constant frictional shear. In Sec. 3
we present a Molecular Dynamic (MD) study of the contact between an elastic block and a rigid cylinder-corrugated substrate.
We present results with and without adhesion, and a qualitative (dimension-based) discussion of the various factors which
influence the contact area. In Sec. 4 we present experiments on adhesion between a human finger and a glass plate,
which are relevant for haptic applications\cite{Haptic}.

\begin {comment}
In this paper we present a molecular dynamics (MD) study of the dependency
of the contact mechanics on the sliding speed when an elastically
soft slab (block) is sliding on a rigid substrate with a ${\rm sin} (q_0 x)$ surface height profile. The atoms on the
block interact with the substrate atoms by Lennard-Jones potentials. We consider contacts with and without adhesion.
For low sliding speeds the contact width is found to be nearly velocity independent, while for high speeds it decreases when
adhesion is included, and increases without adhesion.
We show that when adhesion is included the contact is asymmetric, extending further on the opening crack side then on the closing crack side.
We attribute this to lattice pinning: on the opening crack side the crack tip perform stick-slip motion, where
atoms snap out of contact in rapid events, with atom velocities unrelated to the block driving speed,
followed by ``long'' time periods where the crack tip is pinned. In the rapid
slip events elastic waves (phonons) are emitted from the crack tip resulting in a larger crack propagation energy then the
adiabatic value. This effect is closely related to lattice trapping, the velocity gap
and hysteresis effects observed in model studies of crack propagation in solids\cite{Marder,Per1,Per2,Per3,Per4}.

One particular interesting topic involves the the contact
between a human finger and a glass plate, of importance in haptic applications\cite{Haptic}.
Thus, experiments have shown that the contact area between a human finger and a flat smooth glass surface decreases
as a tangential force is applied to the finger in addition to the normal squeezing force.
This has been interpreted as resulting from the influence of the
tangential force on an adhesive contact\cite{PNAS,PRL}. However, we will show that
there is no macroscopic adhesion between a dry human finger
and a clean glass plate, and we suggest that the decrease in the contact area instead result from large
deformations of a (non-adhesive) complex layered material.
\end{comment}

\vskip 0.3cm
{\bf 2 Theory}

In this section we briefly review theories developed for the variation in the contact area
between an elastic ball and a flat substrate surface when a normal and tangential force is applied to the ball.
We consider both the case when there is no slip in the contact area, and when slip occurs but with constant shear stress.
The former case was considered in a classical study by Savkoor and Briggs\cite{Savkoor}, while the latter case was only very recently
(correctly) studied by Menga, Carbone and Dini\cite{Menga} .

\begin{figure}
\centering
\includegraphics[width=0.45\textwidth]{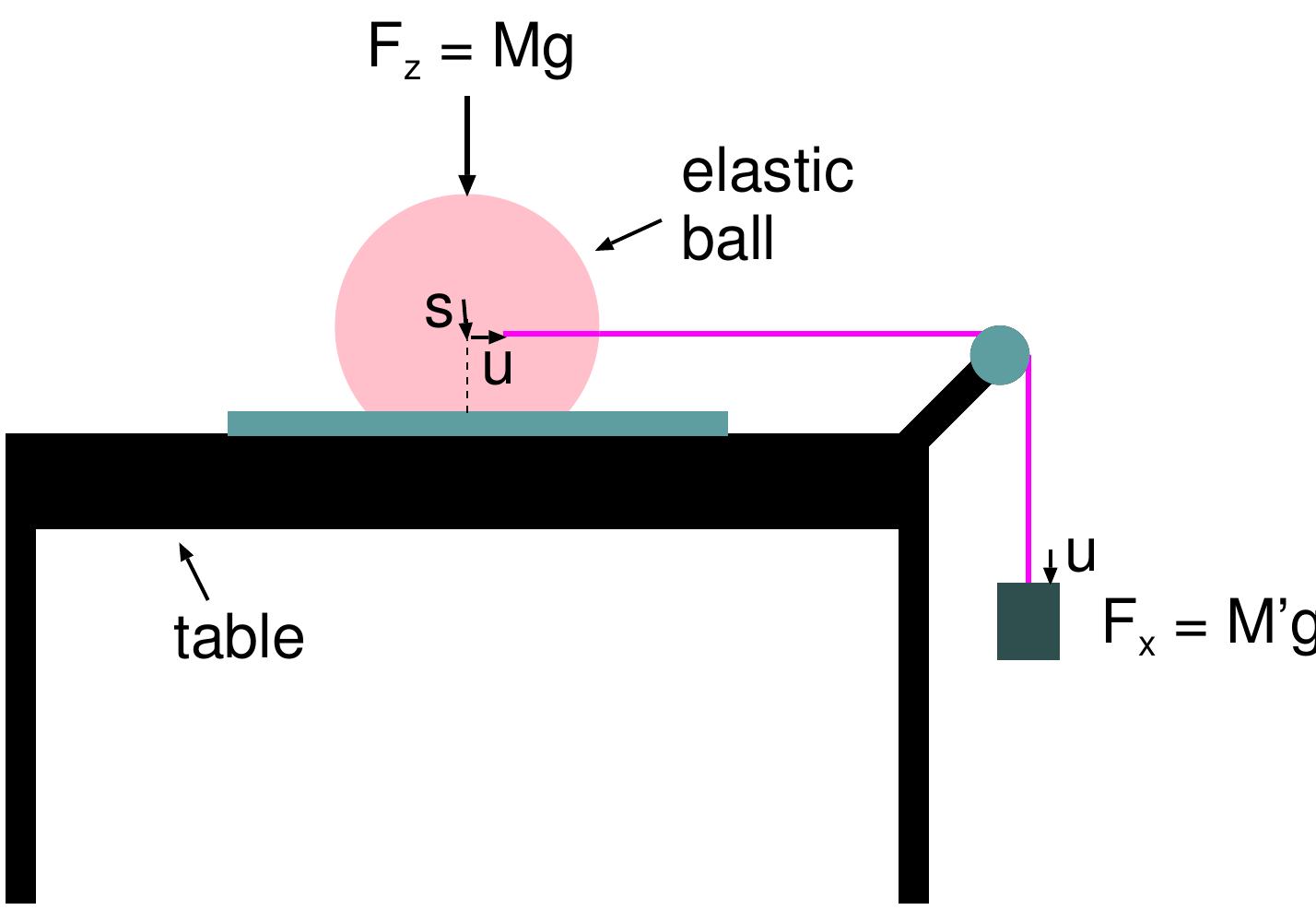}
\caption{\label{BallPlate} An elastic ball squeezed against a flat and rigid substrate. 
The friction force $-F_x$ acting on the ball at the ball-substrate interface is equal in magnitude
to the force $F_x$ from the rope attached to the ball. The center of mass of the ball displaces laterally
a distance $u$ relative to the center of mass of the ball-substrate contact area.}
\end{figure}

Consider an elastic ball (radius $R$) squeezed against a flat counter surface with the force $F_z$.
Assume first that there is no slip between the ball and the counter surface.
The substrate acts on the ball with the tangential (friction) force $-F_x$.
Since the total force acting on the ball must vanish
the force from a rope must equal $F_x$ as indicated in Fig. \ref{BallPlate}.

The total potential energy is
$$U=U_{\rm el}- w_0 \pi r^2 -F_x u -F_z s\eqno(1)$$
The displacements $s$ and $u$ are assumed to vanish before the forces $F_z$ and $F_x$ are applied to the
ball. We assume that the energy $U_{\rm el} =U_1+U_2$ is the sum of the elastic deformation
energies in the ball due to the normal and tangential
forces, respectively. When $F_x$ and $F_z$ are constant, the equilibrium configuration of
the ball-substrate contact is obtained by requiring that the potential energy
vanish to linear order in the variations of the contact radius $r$, and the vertical and horizontal displacement $s$ and $u$. Denoting
$\delta U = U(r+\delta r, s+\delta s, u+\delta u)-U(r,s,u)$ we get
$$\delta \left [ U_{\rm el} -  w_0 \pi r^2 \right ]  -F_x  \delta u -F_z \delta s=0\eqno(2)$$
Since
$$F_x  \delta u +F_z \delta s = \delta W\eqno(3)$$
is the work done by the external force field acting on the sphere, we can write
$$\delta \left [ U_{\rm el} -  w_0 \pi r^2 \right ]= \delta W\eqno(4)$$
This equation is more general than (3) since it is also valid when the external force field is not constant. It states
that the work done by the external forces acting on the ball is equal to the change in the internal energy of the system.

From the JKR theory we know that
$$U_1=E^* \left (rs^2 - {2\over 3} {s r^3\over R} +{1\over 5} {r^5\over R^2}\right ) \eqno(5)$$
The lateral force $F_x$ induces the relative displacement $u$ between the center of mass of the
ball-substrate contact region and the solid far away from the contact region.
We expect the displacement $u$ to be proportional to $F_x$
so that $ku = F_x$, where $k$ is the transverse contact stiffness
which depends on the radius $r$ of the contact and on the tangential stress distribution acting on the ball. The stiffness
$k=k(r)$ can be calculated using the theory of elasticity (see Appendix A).
The contribution to the elastic energy from the tangential stress distribution $\tau({\bf x})$ can be written as
$$U_2 = {1\over 2} \int d^2x \ \tau({\bf x}) u_x({\bf x})$$
where $u_x({\bf x})$ is the lateral displacement of a point ${\bf x}=(x,y)$
on the ball in the contact region relative to the solid far away from the contact region.
If $\tau({\bf x})=\tau$ is constant we get
$$U_2 = {1\over 2} \tau  \int d^2x  \ u_x({\bf x}) =  {1\over 2} \tau (\pi r^2 u)= {1\over 2} u F_x $$
and if instead $u_x=u$ is constant
$$U_2 = {1\over 2} u  \int d^2x \ \tau({\bf x}) = {1\over 2} u F_x $$
Thus in both cases $U_2 = k u^2/2$, but with different stiffness $k$. For the constant shear stress case $k=\alpha E^*r$
and for the no-slip condition $k=\alpha' E^*r$
where $\alpha = (1-\nu)/(A+B\nu)$ and $\alpha' = (1-\nu)/(A'+B'\nu)$, where $\nu$ is the Poisson ratio
and where $A \approx 0.54$, $B\approx -0.27$, $A'=0.5$ and $B'=-0.25$ (see Appendix A).
The total elastic energy:
$$U_{\rm el} =E^* \left (rs^2 - {2 \over 3} {r^3s\over R} +{1\over 5} {r^5\over R^2}\right ) + {1\over 2} k(r)  u^2\eqno(6)$$

Let us now calculate the work $\delta W$. We consider two different cases, namely the ``classical''
case when no slip occur in the contact region and $F_x$ is constant (as expected from
Coulombs friction law if the normal force $F_z$ is constant), and a second case where the
frictional shear stress $\tau$ at the interface is constant, where $F_x = \pi r^2 \tau$ change as the contact area change.
This latter case is expected to hold for smooth soft solids in sliding contact (see below).


\vskip 0.2cm
{\bf 2.1 Constant shear force}

We assume that no slip occurs in the contact area so that $\delta u_x = \delta  u$ is the same
everywhere in the contact area, and that the tangential force $F_x$ is constant.
This case was studied in a classical paper by Savkoor and Briggs\cite{Savkoor}
(see also Ref. \cite{Jon}). Let us consider the work done by the external force field when we change $(r,s, u)$ by
$(\delta r, \delta s, \delta u)$.
The work done is given by (3) $\delta W = F_x \delta  u + F_z \delta s$, and
since $F_x$ is constant
$$\delta W_x = F_x \delta  u = \delta (F_x  u)$$
Using $k  u = F_x$ we get
$$\delta W_x = \delta (k  u^2 )$$
Substituting this in (4) gives
$$\delta \left [ U_1+{1\over 2} k  u^2 -  w_0 \pi r^2 \right ]= F_z\delta s + \delta (k u^2 )$$
or
$$\delta \left [ U_1-{1\over 2} k  u^2 -  w_0 \pi r^2 \right ]= F_z\delta s\eqno(7)$$
Using  that $ u = F_x/k$ we get
$$\delta \left [ U_1-{F_x^2 \over  2 k} -  w_0 \pi r^2 \right ]= F_z\delta s\eqno(8)$$
Using  that $k=\alpha' E^* r$ we get
$$\delta U_1- \left [w_0 2 \pi r -{F_x^2\over 2 \alpha' E^*r^2}\right ] \delta r= F_z\delta s$$
or
$$\delta U_1- 2\pi r \left [w_0 - {F_x^2\over 4\pi \alpha' E^*r^3} \right ] \delta r= F_z\delta s\eqno(9)$$
Hence in this case the tangential force results in an effective decrease in the adhesion. Using (5) and (9)
we obtain the radius $r=r_0$ of the circular contact region
$$r_0^3 = {3 R\over 4E^*} \left (2F_{\rm a}+F_z+2\left [F_{\rm a}F_z+F_{\rm a}^2-{1\over {2 \alpha'}} F_x^2\right ]^{1/2}\right )\eqno(10)$$
where $F_{\rm a} = 3 \pi R w/2$, which is the classical result obtained by Savkoor and Briggs\cite{Savkoor}. When $F_x=0$,
(10) reduces to the standard JKR result. Note also that for small $F_x$ the contact radius (and the contact area) depends linearly on
$F_x^2$, as expected as the contact area must be unchanged as $F_x$ is replaced by $-F_x$. A linear dependency of the contact area on $F_x^2$
has also been observed experimentally before the onset of sliding\cite{PNAS}.

\vskip 0.2cm
{\bf 2.2 Constant shear stress}

When no slip occurs in the contact region, as assumed above, the shear stress will be highly non-uniform with singularities at the
edges of the contact region. In Ref. \cite{Am} it was assumed that the derivation in Sec. 2.1 is valid also when the shear stress
is uniform in the contact region, but with the stiffness $k(r)$ calculated assuming an uniform shear stress rather then uniform displacement.
However, the derivation in Sec. 2.1 is equivalent to the minimization of the total energy assuming
a constant shear force $F_x$. This approach is no longer valid when the shear stress is constant since the shear force $F_x = \tau A$ now depends
on the contact area $A=\pi r^2$, which will vary when the contact radius $r$ varies (B. Persson thank M. Ciavarella for pointing this out).
Still the change in the internal energy (elastic energy + interfacial binding energy)
$U_1$ must equal the work done by the external forces acting on the solid, and this condition was used by
Menga, Carbone and Dini\cite{Menga} to obtain the correct equation for the dependency of
the contact area on the shear force (see also \cite{Kim}).

When we change the radius of the contact area with $\delta r$ the work by the shear force
will have two contributions, namely one from the shear stress in
the area $\delta A$ (annular segment between $r$ and $r+\delta r$)
and one from the original area $A$:
$$\delta W_x=
\tau \int_{\delta A}  d^2x \ \delta u_x({\bf x})+\tau \int_{A}  d^2x \ \delta u_x({\bf x})\eqno(11)$$
where we have used that $\tau$ is constant.
Here $u_x({\bf x})$ is the lateral displacement of a material point on the surface relative to the solid far away from the
contact region.
Since $u_x({\bf x})$ is of order $\delta r$ in the
area $\delta A$ the first integral in (11) is of order $(\delta r)^2$ and can be neglected.
Thus, to linear order in $\delta r$
$$\delta W_x /\tau = \int_A d^2x \ \delta u_x({\bf x})$$
$$= \delta \int_A d^2x \ u_x({\bf x}) - 2 \pi r \delta r  \bar u_x $$
$$=  \delta (\pi r^2  u)  - 2 \pi r\delta r  \bar u_x\eqno(12)$$
where
$$\bar u_x = {1\over 2 \pi} \int_0^{2 \pi} d\phi \ u_x(r {\rm cos}\phi, r {\rm sin}\phi )\eqno(13)$$
Using that $k_x \bar u_x = F_x$ and $k  u = F_x$ with $k=\alpha r E^*$ and $k_x = \beta r E^*$, where
$\alpha$ and $\beta$ only depend on the Poisson ratio $\nu$ (see Appendix A) we get with $F_x = \pi r^2 \tau$ that
$ u = \pi r \tau/ (\alpha E^*)$ and $ u_x = \pi r \tau /(\beta E^*)$. Using these equations in (11) gives
$$\delta W_x= \delta \left ({\pi^2 r^3 \tau^2 \over \alpha E^*} \right )  - {2 \pi^2 r^2 \tau^2 \delta r \over \beta E^*}$$
$$= \delta \left ({\pi^2 r^3 \tau^2 \over \alpha E^*} \right )  - \delta \left ({2 \pi^2 r^3 \tau^2 \over 3 \beta E^*} \right )$$
$$= \delta \left ( {\pi^2 r^3 \tau^2 \over \alpha E^*} \right ) \left (1- {2\alpha\over 3\beta} \right)
= \delta \left ( k  u^2 \right ) \left (1- {2\alpha\over 3\beta} \right)\eqno(14)$$
Using that (see Appendix A) $4 \alpha = 3 \beta$ we get
$$\delta W_x= \delta \left ( {1\over 2} k  u^2 \right )$$

Using (4) and (14) gives
$$\delta \left [ U_{\rm el} -  w_0 \pi r^2 \right ]= F_z\delta s + \delta \left ({1\over 2} k  u^2 \right )$$
or
$$\delta \left [ U_1 -  w_0 \pi r^2 \right ]= F_z\delta s$$
i.e. the contact area when the shear stress is constant is independent on the shear stress.

\begin{figure}
\centering
\includegraphics[width=0.5\textwidth]{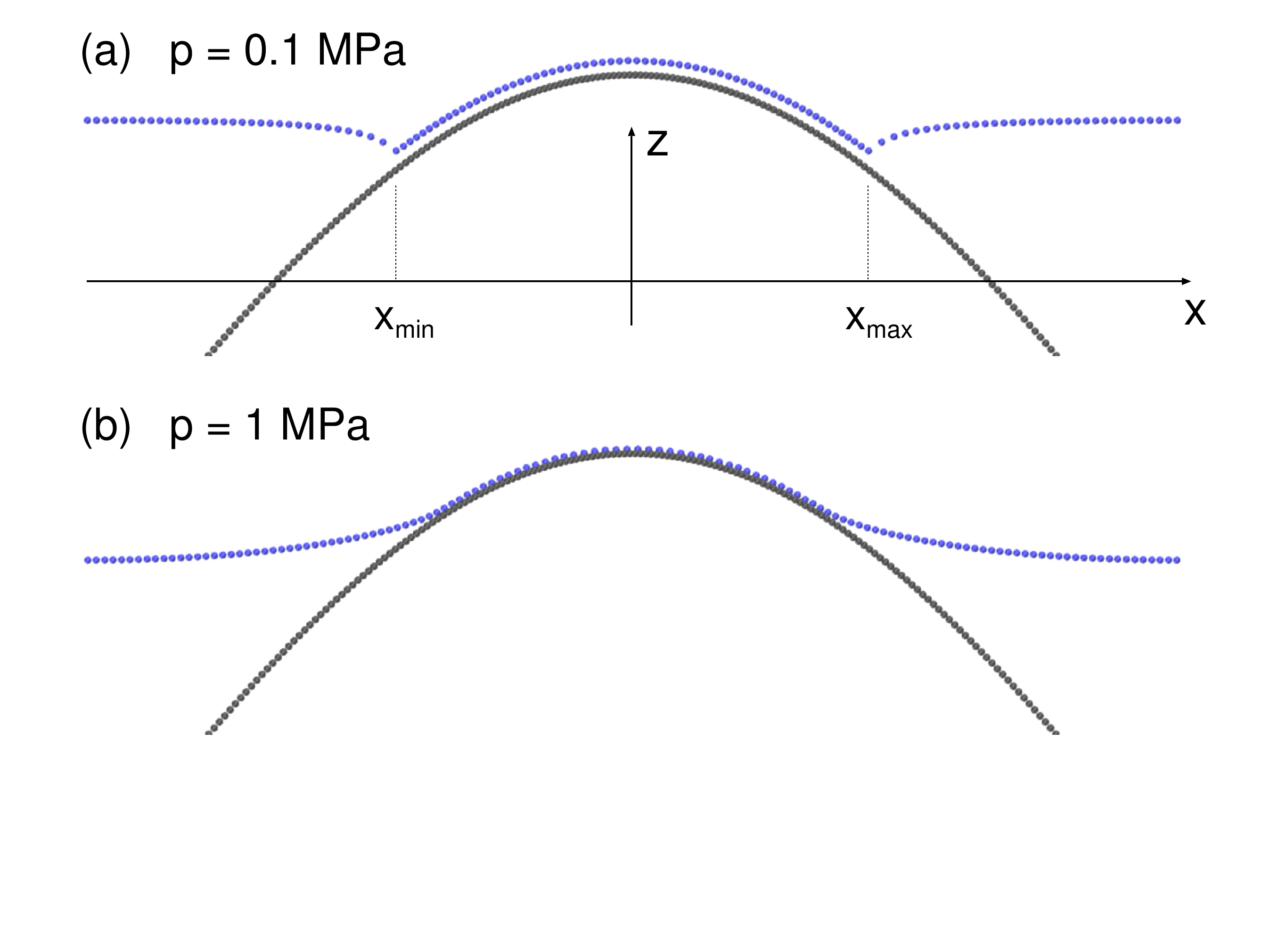}
\caption{\label{ContactPic} The contact area between an elastic slab (block) and a rigid substrate at the temperature $T=0 \ {\rm K}$.
We use periodic boundary conditions in the $xy$-plane with $L_x= 254 \ {\rm \AA}$ and $L_y=14 \ {\rm \AA}$, and the block thickness $L_z \approx 276  \ {\rm \AA}$ (see also main text).The substrate is corrugated with the height coordinate $z=h_0 {\rm sin} (q_0 x)$ ($h_0 = 100  \ {\rm \AA}$ and $q_0 =  \pi /L_x$). We show results (a) with adhesion for the nominal contact pressure $p= F_z /(L_x L_y) = 0.1 \ {\rm MPa}$, and (b) without adhesion for $p=1 \ {\rm MPa}$.}
\end{figure}

\begin{figure}
\centering
\includegraphics[width=0.5\textwidth]{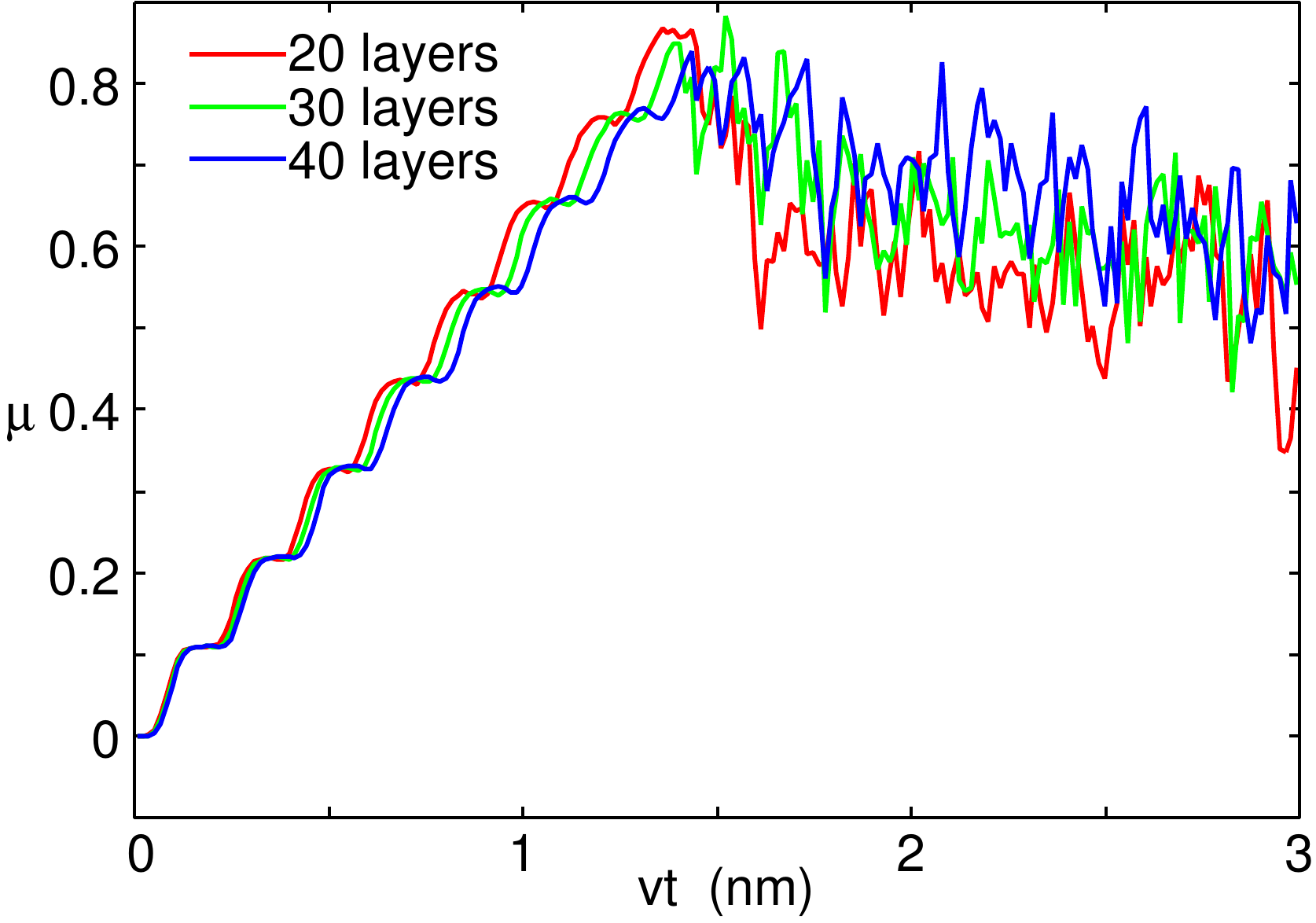}
\caption{\label{fricompavr200} The friction coefficient as a function of displacement distance for elastic slab with $13+7=20$, $23+7=30$ and $33+7=40$ layers.}
\end{figure}

\begin{figure}
\centering
\includegraphics[width=0.5\textwidth]{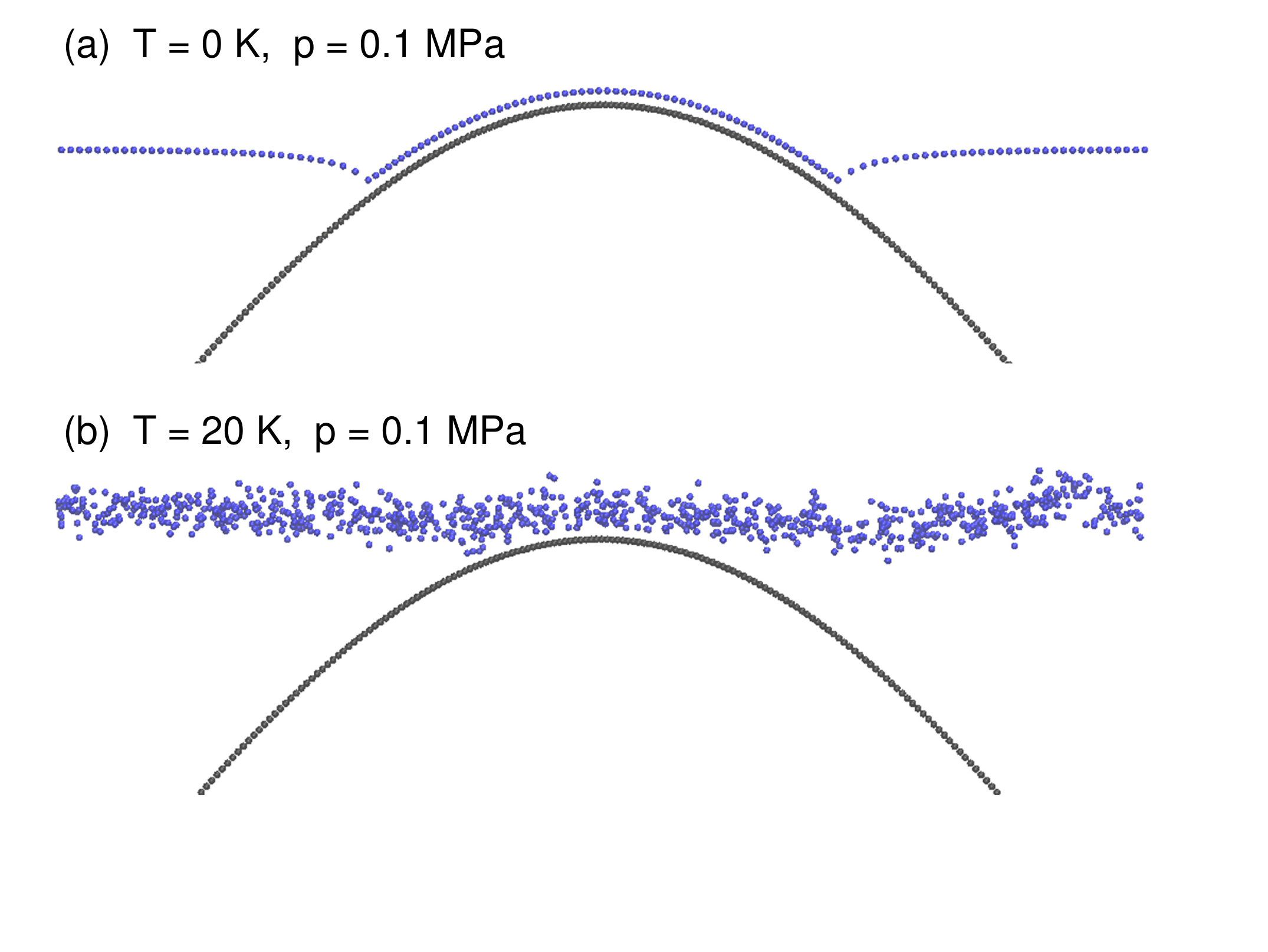}
\caption{\label{T0T20} The contact with adhesion for the nominal contact pressure $p=0.1 \ {\rm MPa}$. (a) at the temperature $T=0 \ {\rm K}$
and (b) for  $T=20 \ {\rm K}$. The contact detach at the higher temperature because this result in an increase in the entropy and a decrease in the system free energy $F=U-TS$.}
\end{figure}

\vskip 0.3cm
{\bf 3. Molecular dynamics simulations}

The two models considered above are very idealized, and in this section we consider a more
detailed model. We find that the contact area is velocity independent for small enough
sliding speeds. We conclude that the velocity dependent contact area observed in Ref. \cite{Chaud}
and \cite{Krick} must result from additional
effects not included in the models studied here, e.g., bulk viscoelasticity, or ``slow'' (thermally activated)
time-dependent relaxation processes at the contacting interface, or mechanical non-linearity (see Sec. 3.3).

\vskip 0.2cm
{\bf 3.1 Model}

Here we present a molecular dynamics (MD) study of the contact between an elastic slab or block (thickness $d$)
with a flat surface squeezed against a rigid substrate with cylinder corrugation (for a related study 
for polymer coated cylinders, see Ref. \cite{mueser1}). 
We consider both contact with and without adhesion. In MD simulations only small system sizes are possible
with linear dimension of order $10-100 \ {\rm nm}$. For the case of no adhesion and stationary contact this is not a major
limitation since in elastic continuum mechanics there is no intrinsic lateral length scale, and the elastic deformation field
scales as $L p/E$ where $L$ is the linear size of the system, and $p/E$ is the only dimension less quantity in the problem.
Here $E$ is the Young's modulus and $p=F_z/L^2$, where $F_z$ is the force squeezing the solids together. Thus the results of the
MD simulations can be re-scaled to correspond to any system size. We note, however, that the friction law acting at the interface
between the solids depends on the interaction potential between the atoms in the MD simulations, but a similar problem
occurs in the (numerical) treatment of the macroscopic system where the result depends on the ``force law" acting between the
grid points at the contact interface.

For adhesive contact the system size has a crucial influence on the contact mechanics. The reason is that for adhesive contact an intrinsic
quantity of dimension length exists, namely $\gamma/E$, where $\gamma$ is the interfacial binding energy (or adiabatic work of adhesion $w_0=\gamma$).
For soft elastic solids like rubber typically $E\approx 1 \ {\rm MPa}$ and $\gamma \approx 0.1 \ {\rm J/m^2}$ giving
$\gamma/E \approx 100 \ {\rm nm}$. If the contact region between two solids is circular-like with the linear dimension $D$, then the surface energy
scales with $D$ like $U_{\rm ad} \sim D^2 \gamma$ and the elastic deformation energy scales as $U_{\rm el} \sim ED^3$.
The ratio $U_{\rm ad}/U_{\rm el} \sim \gamma / (ED)$ decreases as the system size increases. Thus adhesion manifests itself much stronger at short
(say nanometer) length scale than at the macroscopic length scale. Thus, to describe contact mechanics for macroscopic systems in an approximate way using
MD calculations one need to use either a smaller interfacial binding energy or a larger elastic modulus in the MD simulations than what prevails in the macroscopic system.
In the contact mechanics study below the elastic modulus of the block is only $10 \ {\rm MPa}$ (typical for rubber materials)
and we use a very small work of adhesion, $w_0 \approx 0.0027 \ {\rm J/m^2}$, in order to
have partial contact between the solids rather than complete contact which
would prevail for, say, $w_0 =0.1 \ {\rm J/m^2}$.

We consider the contact between an elastic slab and a rigid substrate with the cylinder corrugation (see Fig. \ref{ContactPic})
$$z=h_0 {\rm sin}(q_0 x)$$
where $q_0 = \pi /L_x$ and $0<x<L_x$. We assume periodic boundary conditions in the $xy$ plane with the basic unit having the dimensions
$L_x= 254 \ {\rm \AA}$ and $L_y=14 \ {\rm \AA}$. In order for the contact mechanics not to depend on the block thickness one must choose
the block thickness larger than the diameter of the block-substrate contact region. In the present study the block thickness is
$d \approx 276  \ {\rm \AA}$ unless otherwise stated.

The block is described using the smart-block description described in an earlier publication\cite{Ugo_smart}. In most of the calculations we
use at the interface 13 layers of atoms with the same lattice spacing as in the first layer of block atoms in contact with the substrate.
Above this we use a course-grained description where in each step (total 7 steps) we double the lattice spacing in the
$x$ and $z$ directions (but keep it unchanged in the $y$ direction) and increases the mass of the effective
atoms by a factor of 4 in each step so the mass density is unchanged. The springs between the effective atoms have
elongation and bending stiffness chosen to reproduce the Young's modulus and shear modulus
$G$ specified as input for the calculations. The total thickness of
the block is $d \approx 276  \ {\rm \AA}$ in most of the simulations. However,
we also did some test calculations where we increased the number of atomic layers with
the same lattice spacing as the first layer. Thus, instead on 13+7=20 layers,
we also used 23+7=30 and 33+7=40 layers.

We note that for static contact the smart block description gives
basically the same result as an exact calculation where
the lattice constant is everywhere the same as at the surface (and assuming the same thickness of the block).
However, during sliding lattice vibrations (phonons)
are emitted from the contact region, and for a finite system
without internal damping the block will heat up, and after long enough sliding distance the thermal fluctuations will influence the
contact mechanics and the friction force.
Now, short wavelength phonon's cannot propagate deep into the smart block but will be reflected when the phonon wavelength
become similar to the effective smart block lattice spacing. For this reason it is important to
treat exactly a relative thick layer of atoms at the sliding interface, i.e., to use
the true lattice spacing in this layer. The thicker this layer is the smaller influence will the thermal fluctuations,
resulting from emitted phonon's, have on the contact mechanics.

In the present calculation we include a Langevin type of damping force (proportional to the atom relative velocity)
in the equation of motion for the block atoms during the initial contact formation (no sliding).
After we have obtained the initial contact state (at zero temperature) we remove the damping term and consider so short sliding distances
that frictional heating is negligible.

Fig. \ref{ContactPic}
shows the contact between the elastic slab (block) and the rigid substrate at the
temperature $T=0 \ {\rm K}$ before start of sliding. We only show the first layer of atoms of the block and the substrate at the interface.
The substrate and the block have $N_x=206$ and $128$ atoms along a row in the $x$-direction, and $N_y=11$ and $7$ atoms in the $y$-direction,
respectively.
The substrate and block lattice constants $a_{\rm s} = L_x/N_x \approx 1.233 \ {\rm \AA}$
and $a_{\rm b} \approx 1.984 \ {\rm \AA}$, respectively. The ratio $a_{\rm b}/a_{\rm s} \approx 1.609$ is close to the golden mean 1.618 so the contact is ``almost'' incommensurate.
The elastic slab has the Young's modulus $E=10 \ {\rm MPa}$
and the shear modulus $G=3.33 \ {\rm MPa}$, corresponding to the Poisson ratio $\nu \approx 0.5$.

The atoms at the interface between the block and the substrate interact via
the Lennard-Jones (LJ) interaction potential:
$$V(r) = 4 V_0 \left [\left ({r_0\over r}\right )^{12}-\alpha \left ({r_0\over r}\right )^{6}\right ],$$
where $V_0 = 0.04 \ {\rm eV}$.
For the case of adhesion we use $\alpha = 1$ and $r_0=3.28 \ {\rm \AA}$.
For the case of no adhesion we use $\alpha = 0$ and $r_0= 0.94 \ {\rm \AA}$.
This LJ potential for $\alpha = 1$ gives the adiabatic work of adhesion $w_0 = \gamma \approx 0.0027 \ {\rm J/m^2}$.
For the case of no adhesion we used the rather small $r_0 =  0.94  \ {\rm \AA}$ so that the
block atoms approach the substrate close enough that they feel the corrugated atomic potential; only in this case the
sliding friction force is large enough that it can be detected in the MD simulations.
Thus, when $\alpha = 0$ using the same $r_0=3.28 \ {\rm \AA}$
as used in the case of adhesion, result in a ``superlubric'' sliding state with vanishing friction.

In Fig.  \ref{ContactPic}
we show results pictures of the contact after squeezing the solids into contact (no sliding),
with (a) adhesion for the nominal contact pressure $p= F_z /(L_x L_y) = 0.1 \ {\rm MPa}$, and (b) without adhesion
for $p=1 \ {\rm MPa}$.
We define $x_{\rm max}>0$ and $x_{\rm min}<0$ as the positions of the
leading and receding edge of the contact area, respectively, defined as the $x$-coordinate where the interfacial separation is
equal to $1.2$ times the surface separation at $x=0$. For the case of adhesion, during sliding we can
interpret $x_{\rm max}$ and $x_{\rm min}$ as the positions of opening and closing cracks, respectively.

Fig. \ref{fricompavr200}
shows the friction coefficient as a function of displacement distance $vt$ (with $v=0.1 \ {\rm m/s}$) for the
elastic slab with $13+7=20$, $23+7=30$ and $33+7=40$ layers.
The calculation includes adhesion.
Note that the static or breakloose friction coefficient is $\approx 0.8$ in all three cases, and that the slope
of the $\mu$-distance curve decreases slightly with increasing block thickness as expected from the relation $F_x=Gu/d$,
where $F_x$ is the shear force
and $u/d$ the shear strain. The step-like changes observed during loading in the $\mu$-distance curve
reflect atomic slip processes at the interface where the whole contact region moves
slightly so that the contact becomes asymmetric $x_{\rm max} > |x_{\rm min}|$. Note also that before the onset of macroscopic slip,
the $\mu$-distance curves exhibit very small noise, while during sliding the noise is much larger and increases with increasing sliding distance.
This is due to the elastic waves (phonons) emitted from the sliding contact which are reflected somewhere
in the smartblock region depending on the phonon wavelength; these phonons
perturb the motion of the surface atoms in the contact region and
generate strong fluctuations in the kinetic friction force. The noise in the friction force
decreases as the slab thickness increases by adding 10 and 20 atomic layers to the block thickness;
this result from the reduced number-density of phonons as the block thickness increases.

Fig. \ref{T0T20}  shows the contact with adhesion at the temperature $T=0 \ {\rm K}$ (a)
and for  $T=20 \ {\rm K}$ (b). The contact detaches at the higher temperature because this results in an increase
in the disorder (entropy) and a decrease in the system free energy $F=U-TS$. This occur already for $T=20 \ {\rm K}$ because of the low
interfacial binding energy $\gamma$. This is very similar to the thermal unbinding of membranes\cite{membrane} or structural
phase transitions in adsorbate layers driven by the difference in the vibrational entropy between
different binding sites\cite{entropy}. See also Ref. \cite{mueser2}. 


\begin{figure}
\centering
\includegraphics[width=0.45\textwidth]{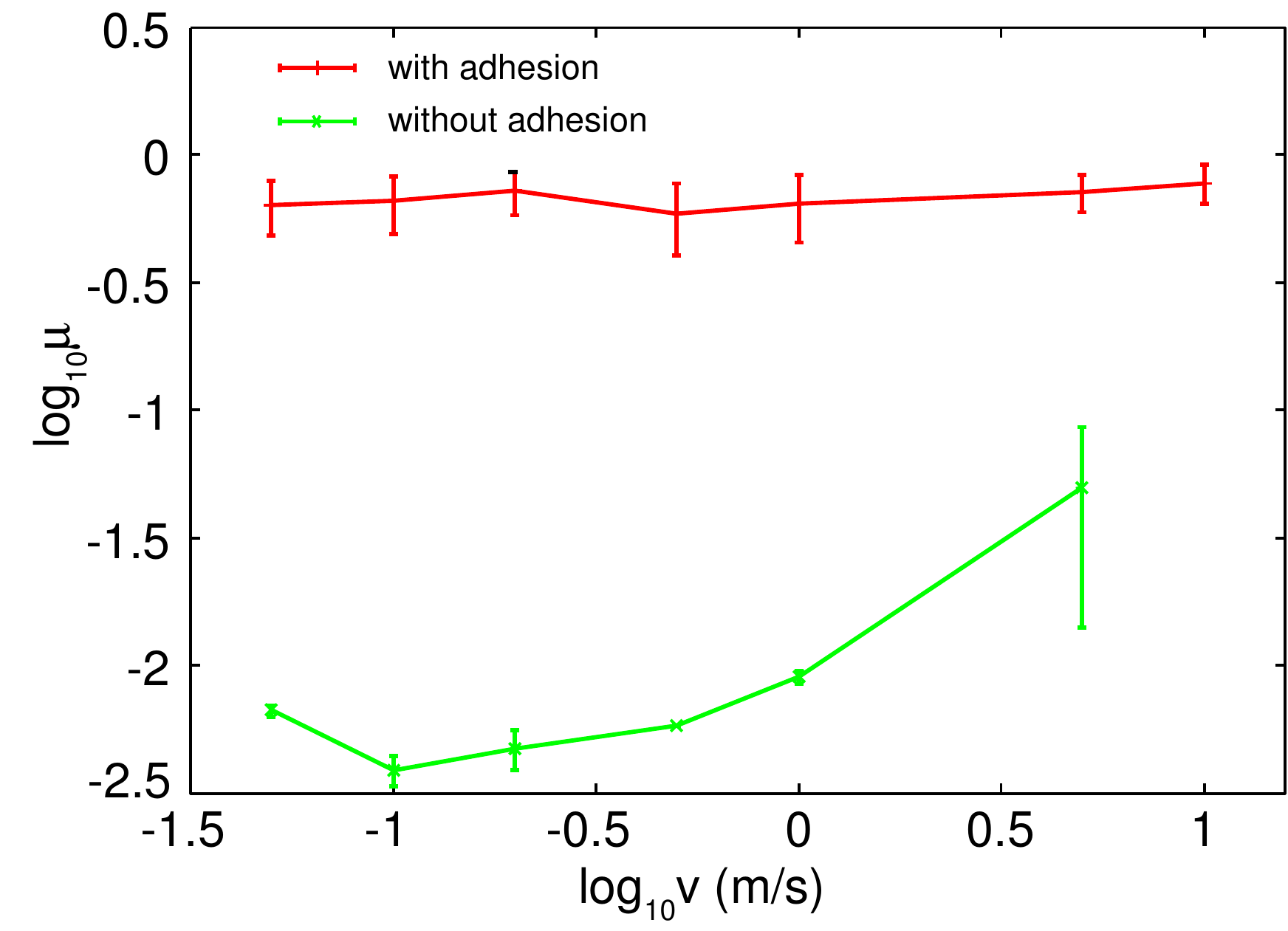}
\caption{\label{friction}
The friction coefficient $\mu = F_x/F_z$ as a function of the logarithm of the sliding speed for contact with adhesion (red curve)
and without adhesion (green curve).}
\end{figure}

\begin{figure}
\centering
\includegraphics[width=0.45\textwidth]{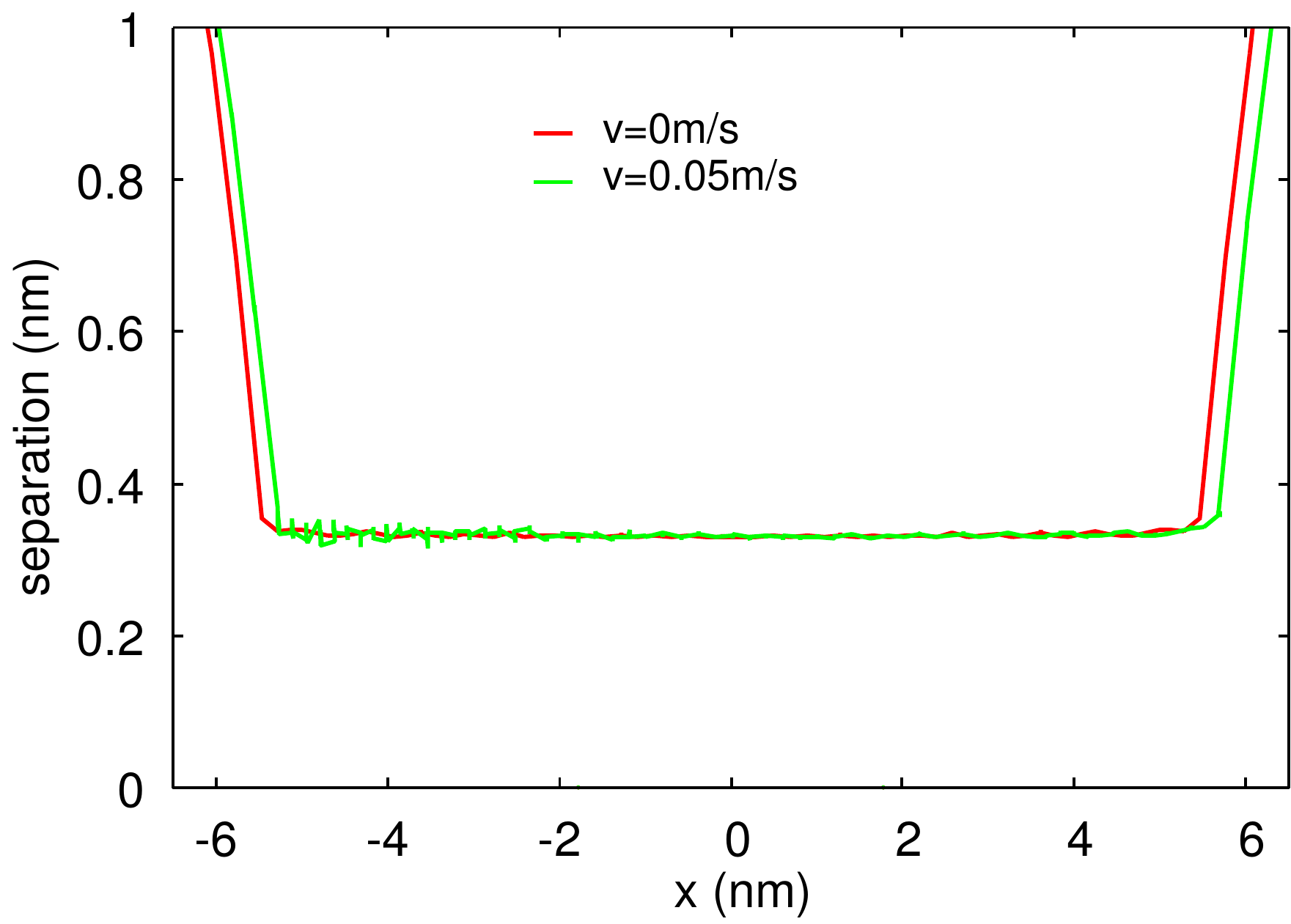}
\caption{\label{separationadhesion}
The interfacial separation (including adhesion) as a function of the lateral coordinate $x$
for stationary contact ($v=0$) (red curve) and for sliding contact $v=0.05 \ {\rm m/s}$ (green).
Note that the for sliding contact the contact becomes slightly asymmetric
(as also shown in Fig. \ref{ratioxmaxxmin}).}
\end{figure}

\begin{figure}
\centering
\includegraphics[width=0.45\textwidth]{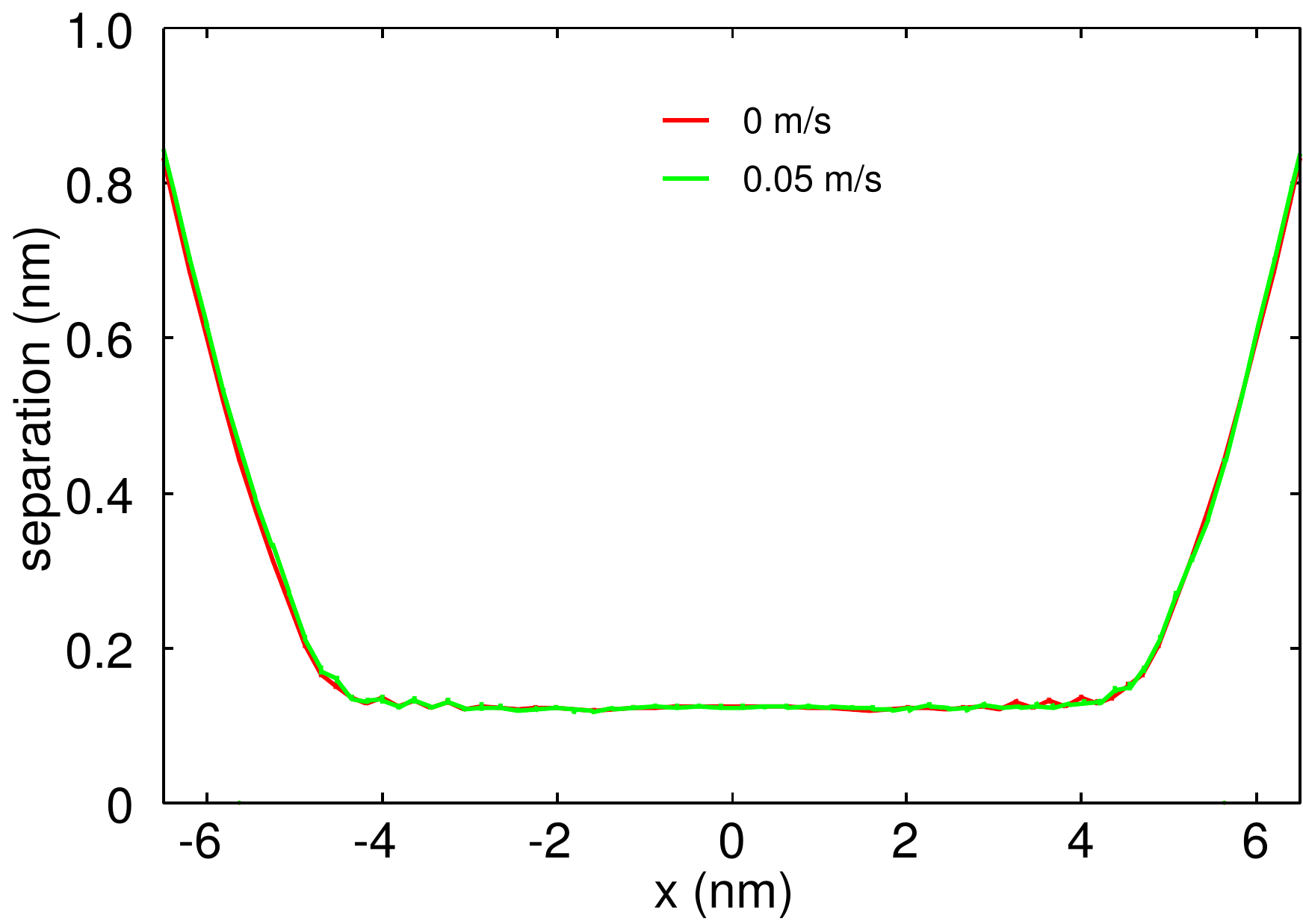}
\caption{\label{separationwithoutadhesion}
The interfacial separation (without adhesion) as a function of the lateral coordinate $x$
for stationary contact ($v=0$) (red curve) and for sliding contact $v=0.05 \ {\rm m/s}$ (green).
At this low sliding speed the friction force is very small $F_x/F_z \approx 0.01$ and there is no
asymmetry in the contact within the noise.}
\end{figure}

\begin{figure}
\centering
\includegraphics[width=0.45\textwidth]{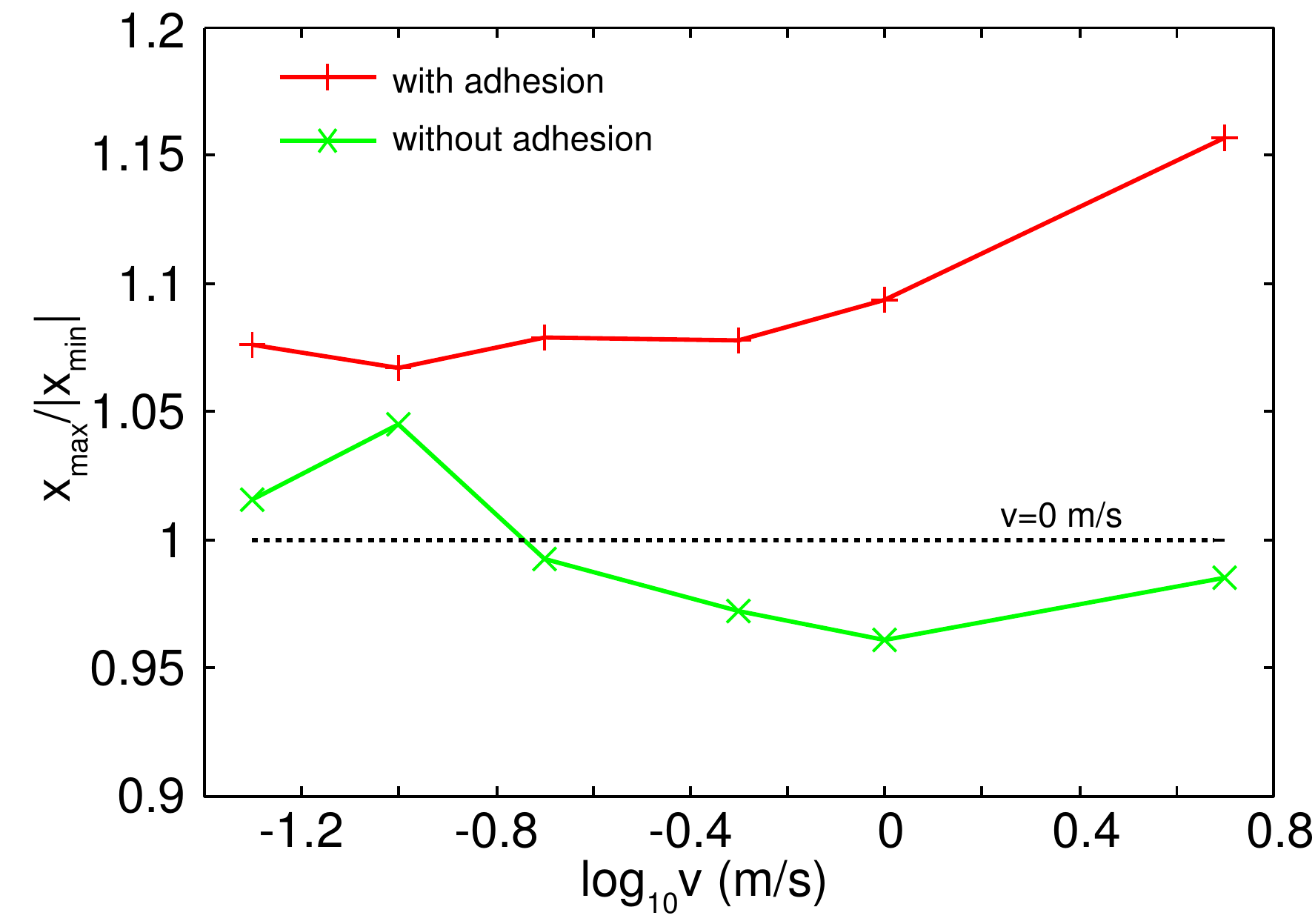}
\caption{\label{ratioxmaxxmin}
The contact area asymmetry factor $x_{\rm max}/|x_{\rm min}|$ as a function of the logarithm of the sliding speed with (red)
and without (green) adhesion.}
\end{figure}


\begin{figure}
\centering
\includegraphics[width=0.45\textwidth]{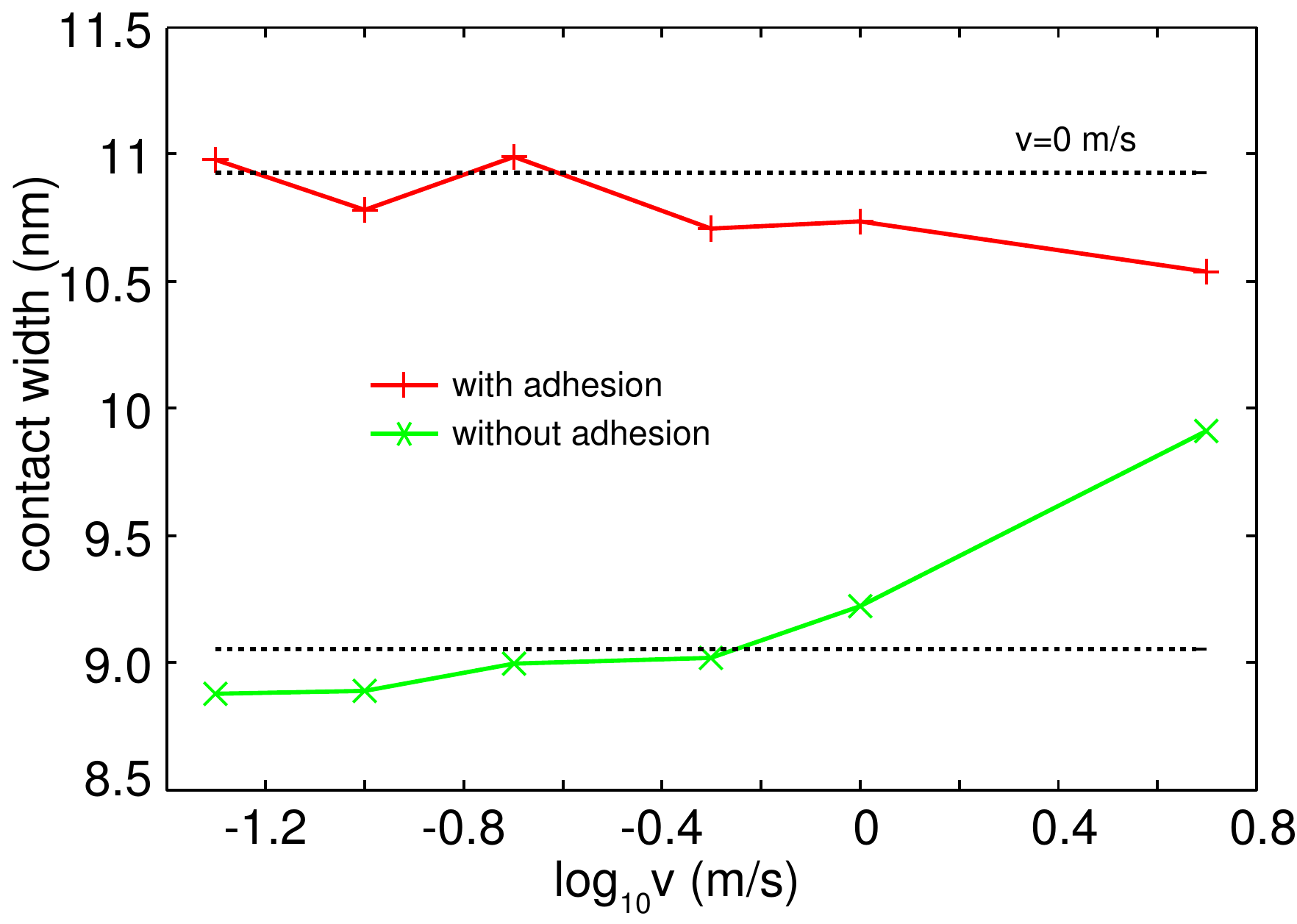}
\caption{\label{contactwidth1}
The geometrical width $w_x=x_{\rm max}-x_{\rm min}$ of the contact area as a function of the logarithm of the sliding speed.}
\end{figure}

\begin{figure}
\centering
\includegraphics[width=0.45\textwidth]{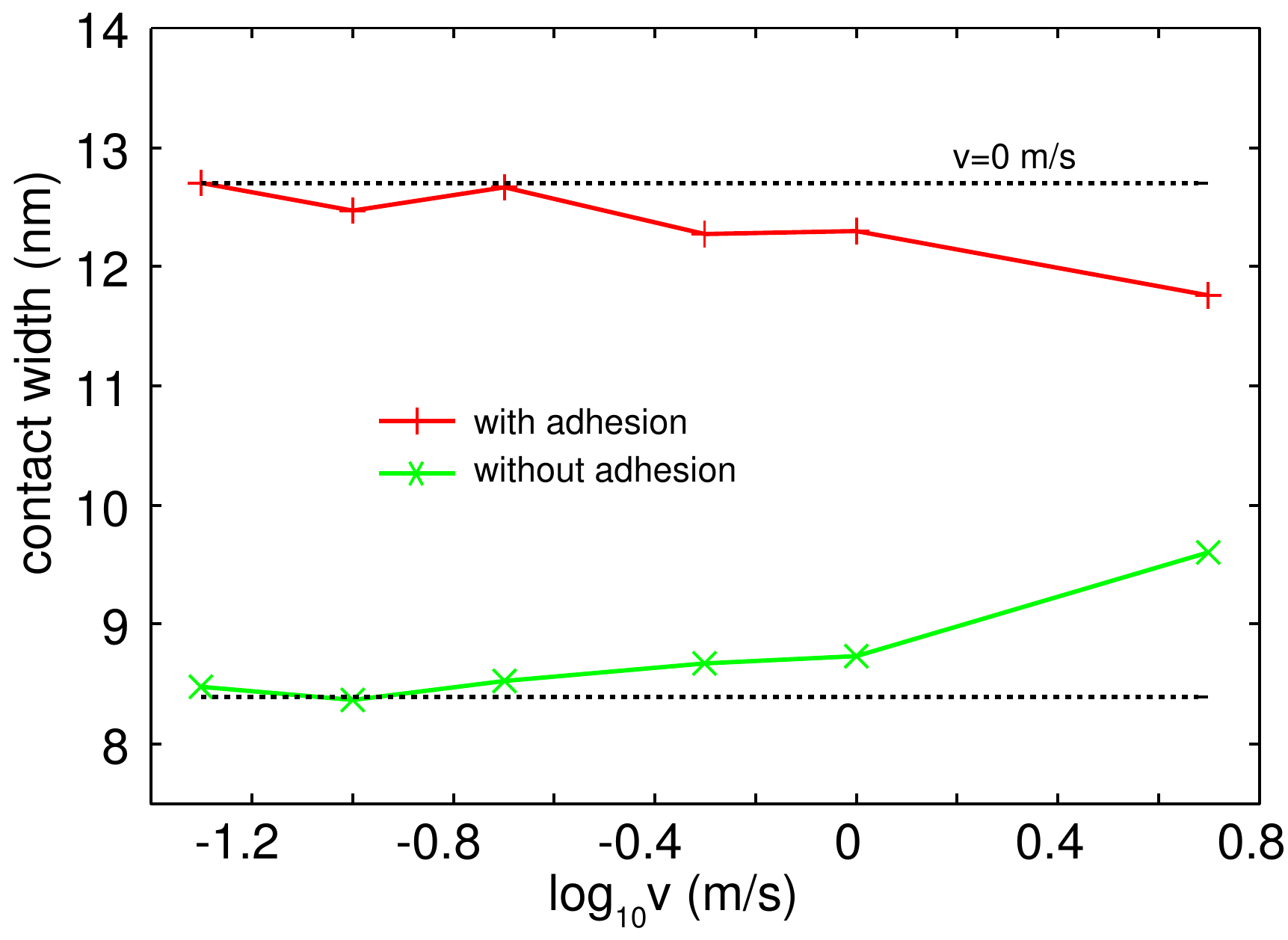}
\caption{\label{contactwidth}
The atom-number width $w_n=N_{\rm b} a_{\rm b}$ of the contact area as a function of the logarithm of the sliding speed.
Here $N_{\rm b}$ is the number of block atoms in contact with the substrate.}
\end{figure}





\begin{figure}
\centering
\includegraphics[width=0.5\textwidth]{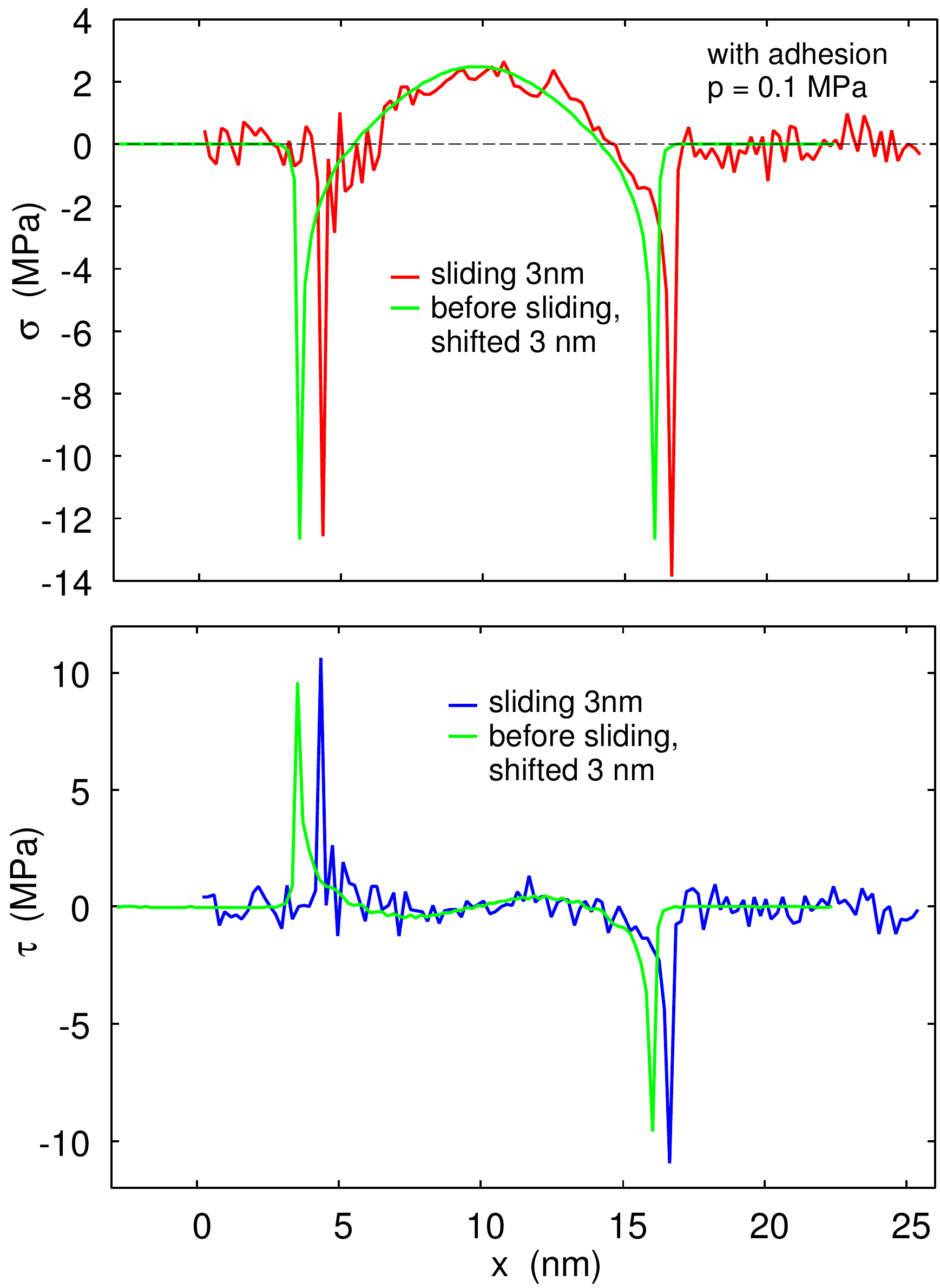}
\caption{\label{1x2sigadhesion}
(a) The perpendicular stress $\sigma = \sigma_{zz}$  and (b) the tangential stress $\tau = \sigma_{xz}$
acting on the block as a function of the spatial coordinate $x$. For the case of adhesion with
the nominal contact pressure $p=0.1 \ {\rm MPa}$ and the sliding speed $v=0.1 \ {\rm m/s}$.
Note that the stress at the opening crack is larger than at the closing crack which is due to lattice pinning effects (see text).}
\end{figure}



\begin{figure}
\centering
\includegraphics[width=0.5\textwidth]{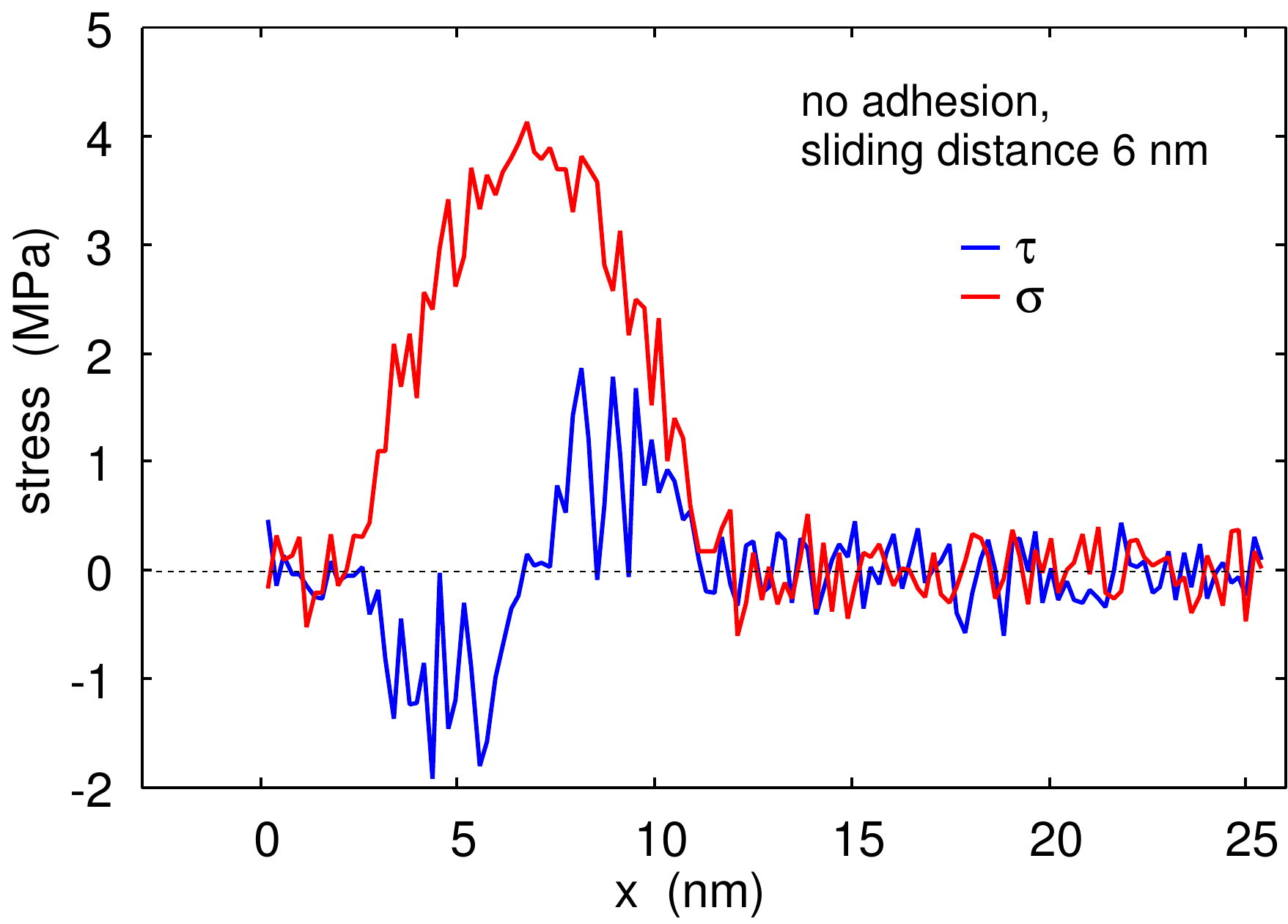}
\caption{\label{1x2sigandtaunoadhesion}
(a) The perpendicular stress $\sigma = \sigma_{zz}$  and (b) the tangential stress $\tau = \sigma_{xz}$
acting on the block as a function of the spatial coordinate $x$. For the case of no adhesion with
the nominal contact pressure $p=1 \ {\rm MPa}$ and the sliding speed $v=0.1 \ {\rm m/s}$. }
\end{figure}

\begin{figure}
\centering
\includegraphics[width=0.3\textwidth]{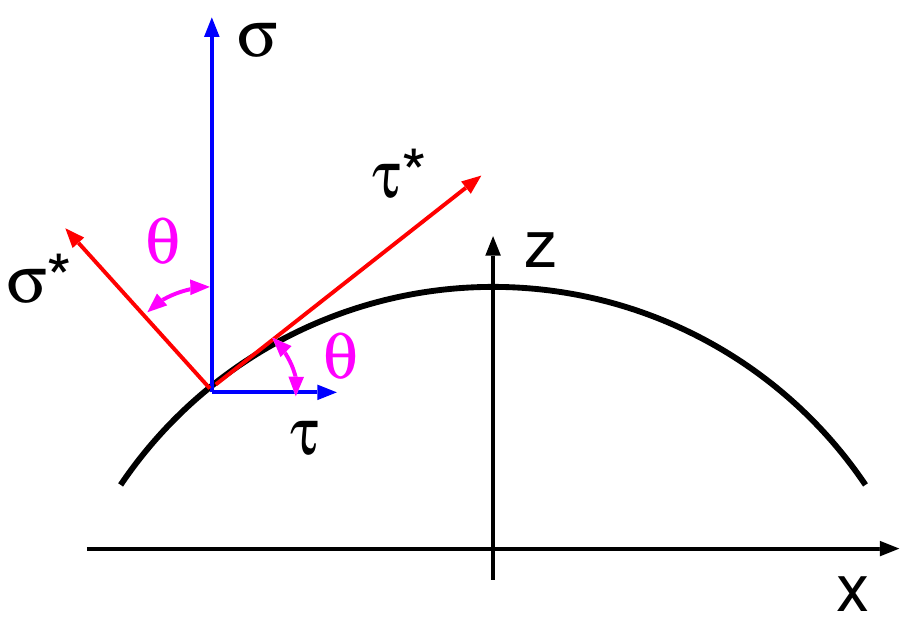}
\caption{\label{Angles}
The stresses $\sigma^*$ and $\tau^*$ normal and tangential
to the substrate profile can be related to the stresses $\sigma = \sigma_{zz}$ and $\tau = \sigma_{xz}$ via
$\sigma^* = \sigma {\rm cos}\theta -\tau {\rm sin}\theta$ and $\tau^*  = \sigma {\rm sin}\theta +\tau {\rm cos}\theta$,
where ${\rm tan}\theta = z'(x)=q_0h_0 {\rm cos} (q_0 x)$ is the slope of the substrate profile.
}
\end{figure}

\begin{figure}
\centering
\includegraphics[width=0.5\textwidth]{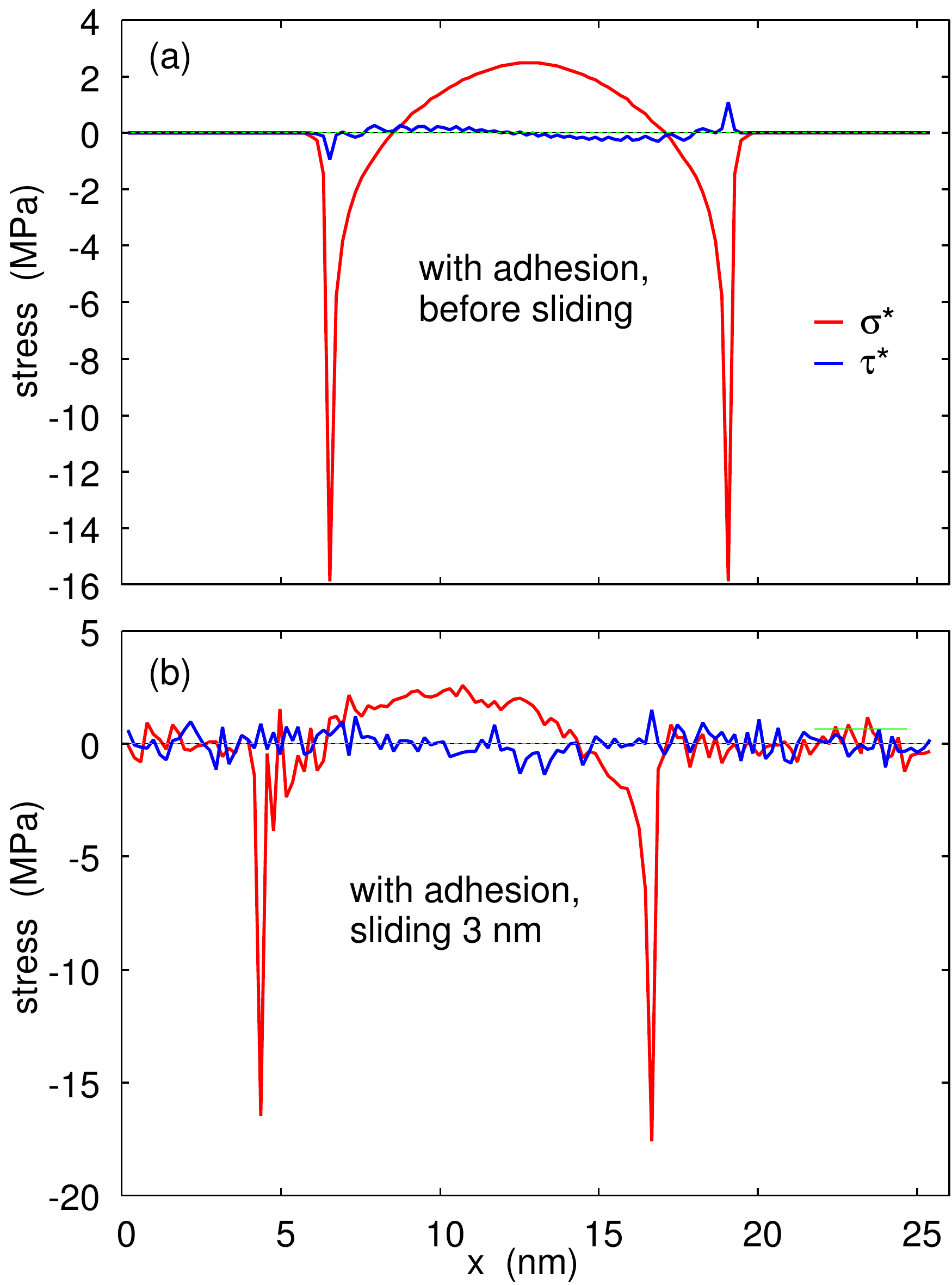}
\caption{\label{1x2sigmaStaadhesion}
The normal stress $\sigma^*$  (red line) and the shear stress $\tau^*$ (blue)
acting on the block as a function of the spatial coordinate $x$.
(a) Only squeezing and (b) after sliding $3 \ {\rm nm}$ at the sliding speed $v=0.1 \ {\rm m/s}$.
The nominal contact pressure $p=0.1 \ {\rm MPa}$.
}
\end{figure}

\begin{figure}
\centering
\includegraphics[width=0.5\textwidth]{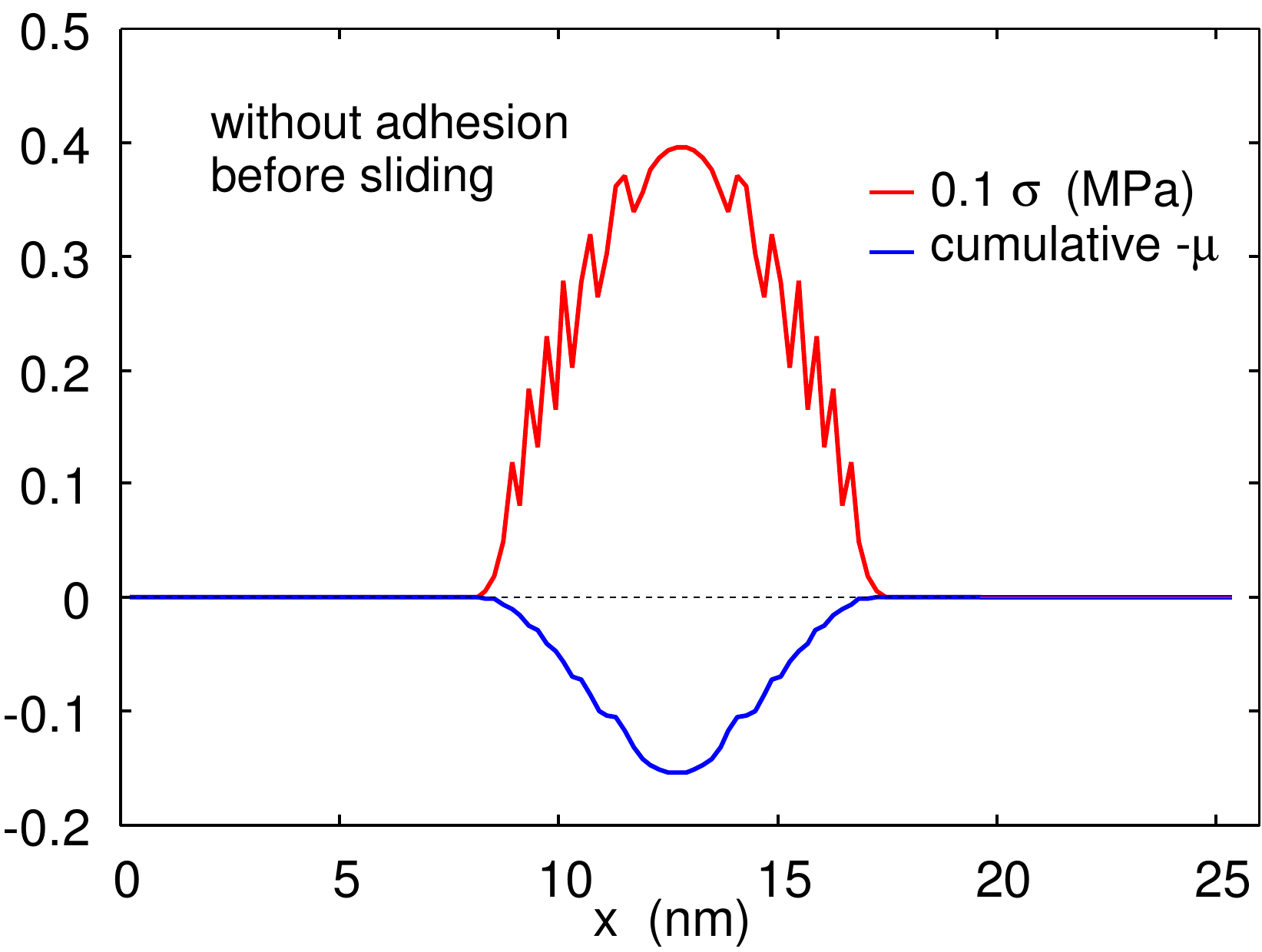}
\caption{\label{1x2cumulativemuNoAdhesionsqueeze}
The contact pressure $\sigma$ (in MPa scaled by a factor 0.1) and the cumulative
normalized (negative) friction coefficient $-\mu(x)=F_x(x)/F_z$, as a function of the position $x$.
Here $F_x(x_1)$ is the force acting on the sliding block from the substrate including
only the tangetial stress in the region $0<x<x_1$.
The results are without adhesion after squeezing the solids into contact
but no lateral sliding.
}
\end{figure}

\begin{figure}
\centering
\includegraphics[width=0.5\textwidth]{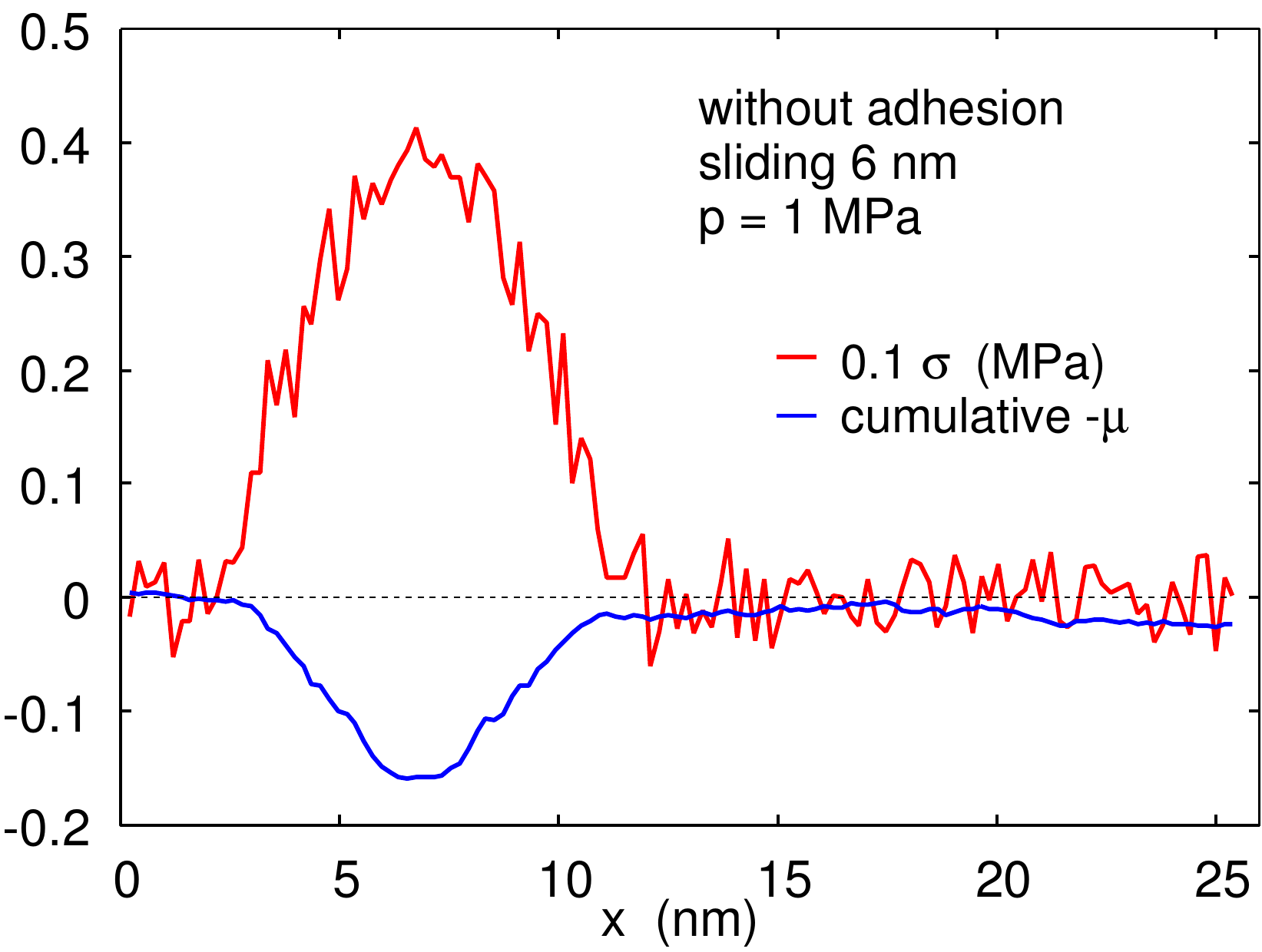}
\caption{\label{1x2cumulativemuNoAdhesion}
The contact pressure $\sigma$ (in MPa scaled by a factor 0.1) and the cumulative
normalized (negative) friction coefficient $-\mu(x)=F_x(x)/F_z$ as a function of the position $x$.
Here $F_x(x_1)$ is the force acting on the sliding block from the substrate including
only the tangential stress in the region $0<x<x_1$.
The results are without adhesion after $6 \ {\rm nm}$ lateral sliding.
}
\end{figure}

\begin{figure}
\centering
\includegraphics[width=0.5\textwidth]{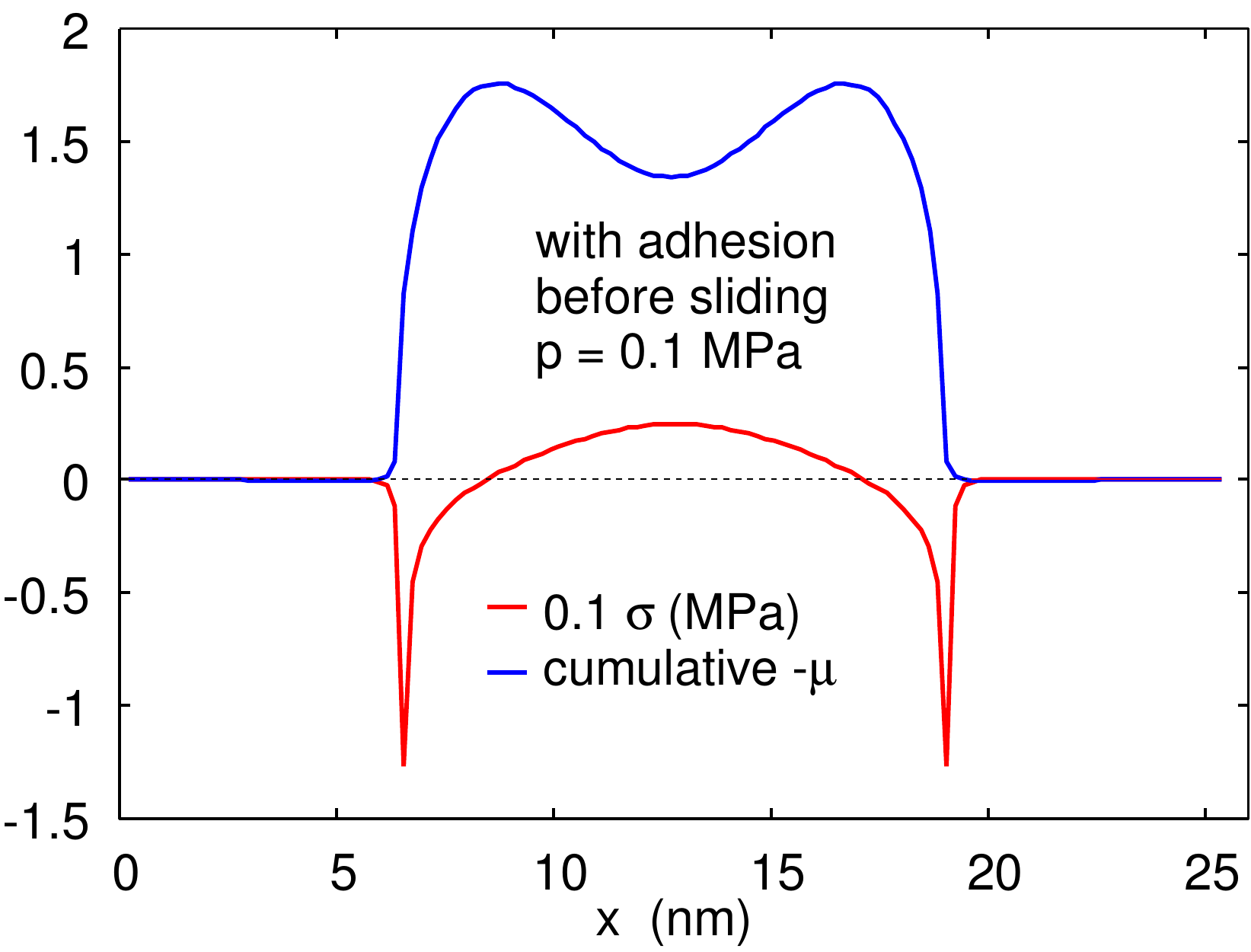}
\caption{\label{1x2cumulativemuWithAdhesionsqueeze}
The contact pressure $\sigma$ (in MPa scaled by a factor 0.1) and the cumulative
normalized (negative) friction coefficient $-\mu(x)=F_x(x)/F_z$ as a function of the position $x$.
Here $F_x(x_1)$ is the force acting on the sliding block from the substrate including
only the tangential stress in the region $0<x<x_1$.
The results are with adhesion after squeezing the solids into contact
but no lateral sliding.
}
\end{figure}

\begin{figure}
\centering
\includegraphics[width=0.5\textwidth]{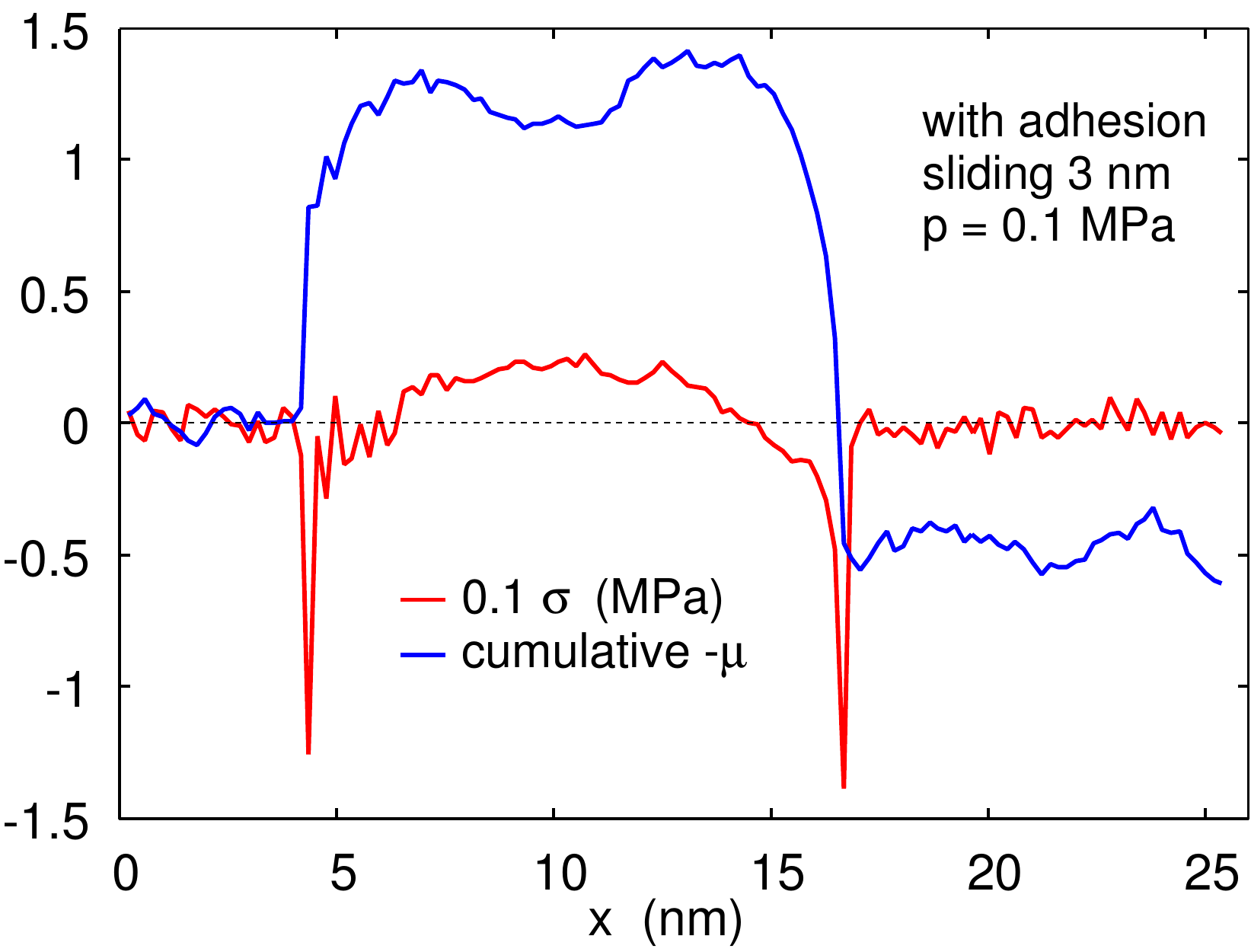}
\caption{\label{1x2cumulativemuWithAdhesion}
The contact pressure $\sigma$ (in MPa scaled by a factor 0.1) and the cumulative
normalized (negative) friction coefficient $-\mu(x)=F_x(x)/F_z$ as a function of the position $x$.
Here $F_x(x_1)$ is the force acting on the sliding block from the substrate including
only the tangential stress in the region $0<x<x_1$.
The results are with adhesion after $3 \ {\rm nm}$ lateral sliding.
}
\end{figure}

\vskip 0.2cm
{\bf 3.2 Results}

Fig. \ref{friction} shows the kinetic friction coefficient $\mu = F_x/F_z$ as a
function of the logarithm of the sliding speed for contact with adhesion (red curve)
and without adhesion (green curve). The friction coefficients were obtained after sliding $3 \ {\rm nm}$ at the given sliding speeds.
The nominal contact pressure acting on the upper surface of the block is $p=0.1 \ {\rm MPa}$ with adhesion, and  $1.0 \ {\rm MPa}$ without adhesion.
Note that with adhesion the friction coefficient
is nearly velocity independent and equal to $\approx 0.6$. This corresponds to an average frictional shear stress
$\tau \approx \mu p A_0/A \approx 0.15 \ {\rm MPa}$, which accidentally is very close
to the shear stress when PDMS spheres is sliding on smooth glass surfaces.
The nearly velocity independence of the friction force is due to the fact that the
friction is caused by rapid slip events, where the local slip velocity is unrelated to
the driving speed. The local slip events are easily observed at the opening crack tip where atoms jump (snap)
out of contact in very rapid events followed by ``long'' time periods where the tip
is pinned by the corrugated interfacial atomic interaction potential (see movie online in Ref. \cite{movie}).
 During the rapid snap out of
contact elastic waves (phonons) are emitted from the opening crack tip
and this is the main origin of the friction force for the case of adhesion\cite{Ho, commentHu}.
This effect is closely related to lattice trapping, the velocity gap
and hysteresis effects observed in model studies of crack propagation in solids\cite{Marder,Per1,Per2,Per3,Per4}

For the case of no adhesion the friction coefficient increases for sliding speed above $\sim 1 \ {\rm m/s}$.
We expect this also for the case of adhesion at higher sliding speeds
because it is well known from the theory of cracks that when the crack tip approach the velocity of
elastic wave propagation in the solids (more exactly, the Rayleigh sound speed), the energy needed to propagate the opening crack
diverges\cite{div}. In the present case the sound velocities in the block is of order $100 \ {\rm m/s}$, so on
approaching this velocity we expect the friction force to increase drastically.

The emission of sound waves from the opening crack result in a crack propagation energy which is
larger than the adiabatic value. For the closing crack no such rapid events
are expected to occur and the closing crack propagation energy is smaller than the opening crack propagation energy.
This result in an asymmetric contact where $x_{\rm max} > |x_{\rm min}|$.
This asymmetry is easily observed in pictures of the interfacial separation as a function of the
lateral coordinate $x$, where $x=0$ is at the top of the substrate cylinder
asperity. This is shown in Fig. \ref{separationadhesion}
which shows the interfacial separation (including adhesion) as a function of the lateral coordinate $x$
for stationary contact ($v=0$) (red curve) and for sliding contact $v=0.05 \ {\rm m/s}$ (green).
Note that the for sliding contact the contact becomes slightly asymmetric
(as also shown in Fig. \ref{ratioxmaxxmin}).

In the case of no adhesion we observe negligible contact asymmetry in the studied velocity interval.
This is illustrated in Fig. \ref{separationwithoutadhesion}
which shows the interfacial separation (without adhesion) as a function of the lateral coordinate $x$
for stationary contact ($v=0$) (red curve) and for sliding contact $v=0.05 \ {\rm m/s}$ (green).
At this low sliding speed the friction force is very small $F_x/F_z \approx 0.01$ and there is no
asymmetry in the contact within the noise level. At very high sliding speed $v > 10 \ {\rm m/s}$ the friction force
is larger $F_x/F_z > 0.1$, and the asymmetry is negative i.e. opposite to the case of adhesion (not shown).

Fig. \ref{ratioxmaxxmin}
shows the contact area asymmetry
factor $x_{\rm max}/|x_{\rm min}|$ as a function of the logarithm of the sliding speed with (red)
and without (green) adhesion. We have chosen $x=0$ at the top of the substrate asperity,
and $x_{\rm max}>0$ and $x_{\rm min}<0$ are the positions of the
leading edge and receding edge of the contact area, respectively, defined as the $x$-coordinate where the interfacial separation is
equal to $1.2$ times the surface separation at $x=0$. For the case of adhesion (red line) we can
interpret $x_{\rm max}$ and $x_{\rm min}$ as the positions
of the opening and closing crack edges, respectively.

We now study the dependency of the contact area on the sliding speed. There are two ways to define the contact area.
Thus, Fig. \ref{contactwidth1}
shows the geometrical width (projected contact width) $w_x=x_{\rm max}-x_{\rm min}$ as a function of the logarithm of the sliding speed,
while Fig. \ref{contactwidth} shows what we denote as the atom-number
width $w_n=N_{\rm b} a_{\rm b}$.
Here $N_{\rm b}$ is the number of block atoms (in one row along the $x$-axis) in contact with the substrate,
where an atom is defined to be in contact with the substrate when the surface separation
is smaller than 1.2 times the surface separation for $x=0$.
Note that the contact area during sliding at low sliding speed is nearly the same
as for $v=0$. At sliding speeds $v > 0.25 \ {\rm m/s}$ the contact area with adhesion decreases and the contact area without adhesion increases.

Note that the contact atom-number width is larger than the geometrical width.
This is due to
the fact that the binding energy between block and substrate atoms tend to accumulate atoms in the contact region.
This costs some additional elastic energy but is compensated by the increased binding energy.

For the contact between a macroscopic rubber ball and a smooth glass surface
experiments have shown that the shear stress during slip is uniform.  The origin of this is due to the following effect:
When a rubber ball with smooth surface is squeezed against a flat surface, the contact pressure
$p$ will usually be of order the rubber Young's modulus $E$ or less, i.e., of order $1 \ {\rm MPa}$ or less.
This contact pressure is much smaller than the adhesive pressure
$p_{\rm ad}$ which is of order $w_0 /d$, where $w_0$ is the work of adhesion and $d$ an atomic distance. Using $w_0=0.1 \ {\rm J/m^2}$ and
$d=1 \ {\rm nm}$ we get $p\approx 100 \ {\rm MPa}$. As long as $p<<p_{\rm ad}$ we
expect the frictional shear stress in the area of real contact to be independent of the contact pressure, i.e., only a function of
the sliding speed. This is in agreement with experiments\cite{France} and computer simulations\cite{Ion}.

However, in our simulations we observe a non-uniform shear stress. This is due to the fact that we use a very small work of adhesion
$w=0.0027 \ {\rm J/m^2}$, so that $w_0/d \approx 2 \ {\rm MPa}$ while the contact pressure
reaches $\approx 10 \ {\rm MPa}$ at the opening and closing
crack tips (see Fig. \ref{1x2sigadhesion}). In addition, inside the contact the atomic positions of the
block are nearly incommensurate with respect to
the substrate atoms so the main frictional interactions occur at the opening and closing crack tips.
This differs from the case of rubber on a glass surface where
the rubber chains at the interface rearrange themselves in the substrate potential and pin the surfaces together.
During sliding nanometer sized pinned regions
undergo stick-slip type of motion, resulting in a frictional shear stress which is nearly uniform in the contact region.

To illustrate the non-uniform nature of the stress at the interface, in Fig.
\ref{1x2sigadhesion} we show
(a) the perpendicular stress $\sigma = \sigma_{zz}$  and (b) the tangential stress $\tau = \sigma_{xz}$
acting on the block as a function of the spatial coordinate $x$, for the case of adhesion with
the nominal contact pressure $p=0.1 \ {\rm MPa}$, and the sliding speed $v=0.1 \ {\rm m/s}$.
Note that the stress is JKR-like but the stress at the opening crack
is larger than at the closing crack which is due to the lattice pinning effects discussed above.

For the contact without adhesion, in Fig. \ref{1x2sigandtaunoadhesion}
we show the perpendicular stress $\sigma = \sigma_{zz}$  and the tangential stress $\tau = \sigma_{xz}$
acting on the substrate as a function of the spatial coordinate $x$. In this case
the nominal contact pressure $p=1 \ {\rm MPa}$ and the sliding speed $v=0.1 \ {\rm m/s}$.
Note that the perpendicular pressure is Hertz-like and the tangential stress highly non-uniform.

The stress $\sigma = \sigma_{zz}$ in Fig. \ref{1x2sigadhesion}(a) and $\tau=\sigma_{xz}$ in Fig. \ref{1x2sigadhesion}(b)
are the stresses along the $z$ and $x$-axis,
i.e., not the stress $\sigma^*$ normal and $\tau^*$ tangential to the (rigid) substrate profile.
However, the latter stresses can be easily obtained from linear combination of $\sigma$ and $\tau$ (see Fig. \ref{Angles}):
$\sigma^* = \sigma {\rm cos}\theta -\tau {\rm sin}\theta$ and $\tau^*  = \sigma {\rm sin}\theta +\tau {\rm cos}\theta$,
where ${\rm tan}\theta = z'(x)=q_0h_0 {\rm cos} (q_0 x)$ is the slope of the substrate profile.

For the case of adhesion,
the stresses $\sigma^*$ and $\tau^*$ acting on the block are shown in
Fig. \ref{1x2sigmaStaadhesion}
as a function of the spatial coordinate $x$.
The nominal contact pressure $p=0.1 \ {\rm MPa}$ and the sliding speed
$v=0.1 \ {\rm m/s}$. We show results (a) after squeezing the solids into contact (zero sliding distance),
and (b) after sliding $3 \ {\rm nm}$. In (b) the average frictional shear stress is only
$\approx 0.15 \ {\rm MPa}$, but the local shear stress is much larger and takes both
positive and negative values.

Fig. \ref{1x2cumulativemuNoAdhesionsqueeze}
shows the contact pressure $\sigma$ (in MPa scaled by a factor 0.1) and the cumulative
normalized (negative) friction coefficient $-\mu(x)=F_x(x)/F_z$, as a function of the position $x$.
Here $F_x(x_1)$ is the force acting on the sliding block from the substrate including
only the tangential stress in the region $0<x<x_1$.
The results are without adhesion after squeezing the solids into contact
but no lateral sliding.
Fig. \ref{1x2cumulativemuNoAdhesion}
shows similar results without adhesion after $6 \ {\rm nm}$ lateral sliding.

Fig. \ref{1x2cumulativemuWithAdhesionsqueeze} and Fig. \ref{1x2cumulativemuWithAdhesion}  shows
results with adhesion after squeezing the solids into contact but no lateral sliding, and after $3 \ {\rm nm}$ lateral sliding, respectively.
Note that only close to the opening crack tip does the cumulative $F_x(x)/F_z$ become negative, i.e., the frictional
shear force is entirely due to processes occurring close to the opening crack tip.

\begin{figure}
\includegraphics[width=0.9\columnwidth]{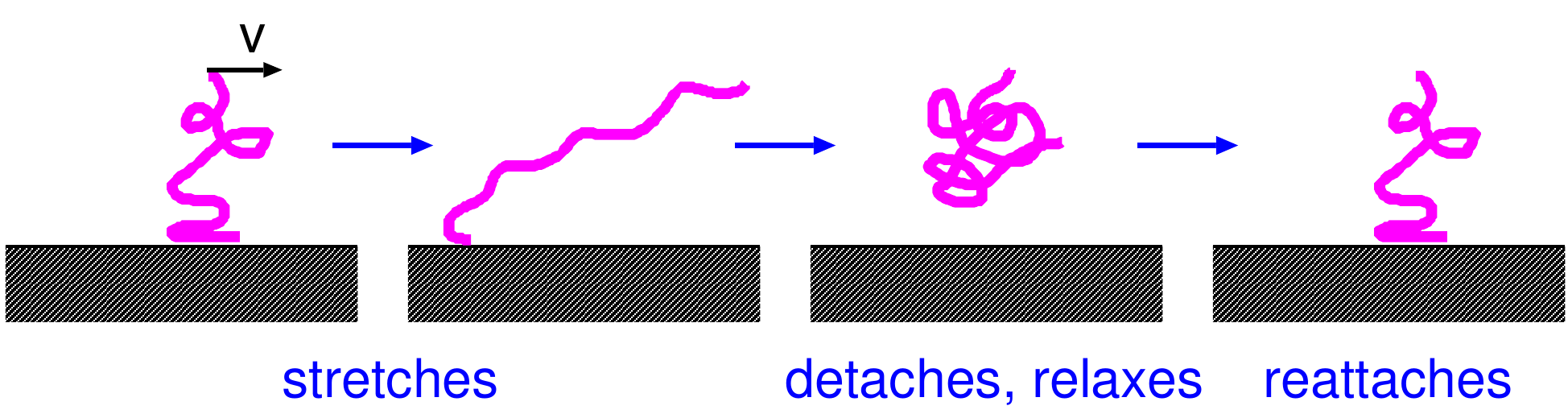}
\caption{\label{PerssonVol}
The classical description of a polymer chain at the rubber-block counter surface interface. During lateral
motion of the rubber block the chain stretches, detaches, relaxes, and reattaches to the surface to repeat the cycle.
The picture is schematic and in reality no detachment in the vertical direction is expected, but only a rearrangement
of molecule segments (in nanometer-sized domains) parallel to the surface from pinned (commensurate-like) to depinned
(incommensurate-like) domains.
}
\end{figure}

\vskip 0.2cm
{\bf 3.3 Discussion}

The most important results of the study above are that the contact area and the friction force are nearly velocity independent for
small velocities ($v < 0.25 \ {\rm m/s}$) in spite of the fact that the shear stress in the contact area is rather non-uniform.
For the case of no adhesion the friction force is very small
because the contribution to the friction force from the surface area where the shear stress is negative is almost completely
compensated by the contribution from the area where the shear stress is positive.
For the case of adhesion the friction force is higher, and is mainly due to energy
dissipation at the opening crack tip, where rapid atomic snap-off events occur
during sliding. This ``edge-dominated friction'' is very different from the frictional
processes expected when a macroscopic rubber sphere is sliding on a substrate.
In the latter case one expects an ``area-dominated friction'' where the shear stress is uniform
within the contact area. In this case the  friction force
arises from stick-slip type of motion of nanometer-sized regions everywhere within the contact region (see Fig. \ref{PerssonVol})\cite{Schall,smooth}.

The results presented above are consistent with a very simple dimensional analysis. Let us consider an elastic sphere of radius $R$
squeezed against a flat rigid surface. Let us first consider the case of no adhesion and zero temperature.
In that case there are only the following quantities in the problem: Elastic modulus $E$, block mass density $\rho$, lattice constant $a$,
radius of curvature $R$ of the sphere, pressure $p=F_z/R^2$ (where $F_z$ is the applied normal force),
and the sliding speed $v$. From these quantities we can construct the
following dimensionless quantities:
$p/E$, $v/c$ (where $c=(E/\rho)^{1/2}$ is the sound velocity) and $a/R$. We assume $a/R <<1$ so only the $a/R \rightarrow 0$
limit interest us, and $a/R$ drop out from the analysis.
In this case the contact area can be written as
$$A = R^2 f(p/E,v/c).$$
The friction coefficient is also a function of only $p/E$ and $v/c$.
For small enough $v/c$ we can expand the contact area $A$ to leasing order in $v/c$ to get
$$A \approx R^2 f(p/E,0)\left [1+g(p/E) \left ( {v\over c}\right )^2\right ]$$
where we have used that the contact area cannot change when the velocity $v=v_x$ change sign.
For a given normal force $p/E$ is a fixed number, and for rubber-materials $p/E$ is typically of order $0.1-1$.
If $p/E$ is of order unity we expect the function $g(p/E)$ to be of order unity and in this case
for $v/c << 1$ the contact area has a negligible dependency on the sliding speed.

When adhesion is included a new dimensional less parameter occur, namely $\gamma /(ER)$ and $A$ will depend on this quantity,
but the velocity dependency will still enter via the dimensionless parameter $v/c$. If $v/c$ is small enough
we can still expand the contact area $A$
to leading order in $v/c$, but now the function $g(p/E)$ is replaced by $g(p/E, \gamma /(ER))$. If both $p/E$  and $\gamma /(ER)$
are of order unity we conclude again that for $v/c << 1$ the contact area has a negligible dependency on the sliding speed.
However, if $\gamma /(ER) << 1$ then the expansion to leading order in $v/c$ will hold only if
$g(p/E, \gamma /(ER)) (v/c)^2 << 1$. But if
$\gamma /(ER) << 1$ and $p/E$ is of order unity adhesion is not very important for the contact area,
so in a typical case we conclude that if  $p/E$ is of order unity,
if $v/c << 1$ the contact area has a negligible dependency on the sliding speed also when adhesion occur.

At non-zero temperatures one more dimensionless parameter enters in the expression for the contact area,
namely $k_{\rm B}T/(E R^3)$. For macroscopic systems this quantity is extremely small. Thus if $E \approx 10 \ {\rm MPa}$
and $R= 1 \ {\rm cm}$ we get at room temperature $k_{\rm B}T/(ER^3) \approx 10^{-21}$. In this case it is no longer possible to
assume $a/R=0$, but we need to include the lattice constant $a$ in the analysis. The parameter $[k_{\rm B}T/(ER^3)][(R/a)^3]
=k_{\rm B}T/(Ea^3)$ is typically of order unity, e.g., using $E=10 \ {\rm MPa}$ and $a=1 \ {\rm nm}$ gives at room temperature
$k_{\rm B}T/(Ea^3) \approx 0.4$. Hence if $p/E$ and $\gamma /(ER)$ are of order unity
we expect a velocity independent contact area for $v/c << 1$,
even when the temperature is nonzero.
Note, however, that when thermal effects are included slow creep motion will occur when the lateral driving
force is small, and the friction force will vanish linearly with the sliding speed as $v\rightarrow 0$.

The adhesion and temperature parameters introduced above may not be the physically most relevant parameters. Thus,
when adhesion is included we expect thermal effects to be important for breaking the bonds at the crack tip
so a more physical motivated dimension less temperature parameter
is $[k_{\rm B}T/(Ea^3)][\gamma /(ER)]^{-1} = k_{\rm B}T R/(\gamma a^3)$.
Similarly, from the JKR theory we expect the contact area to depend on the parameter
$[\gamma /(ER)][E/p] = \gamma /(pR)$.

The situation discussed above changes completely if some new energy dissipation or relaxation process occurs in the problem.
For example, for rubber-like materials a wide distribution
of chain relaxation times enter into the problem. In the simplest case (but not realistic for rubber materials)
a single relaxation time $t^*$ enter as in, e.g.,
the Maxwell or Kelvin-Voigt rheology models. In this case a new ``velocity" $v^* = a/t^*$ can be formed and $\mu$ and $A$ will depend
on $v/v^*$ in addition to the dimension less quantities mentioned above. Since $v^*$ may be much
smaller than the sound velocity $c$, a dependency of $\mu$
and $A$ on the sliding speed may occur already at low sliding speeds.

We note that the binding interaction of the rubber polymer
segments to the substrate may introduce another relaxation time.
Thus, the polymer segments at the rubber surface need some time $t'$ to adjust to the corrugated substrate potential, to
bind as strongly as possible to the substrate after each local slip event.
This defines the velocity $v' = b/t'$, where $b$ is an atomic distance, e.g., the substrate lattice
constant or the length of a polymer segment (bed unit). At the sliding speed $v\approx v'$
the adhesive contribution to the friction will be maximal. Rubber friction studies have shown that typically $v' = 0.001-0.1 \ {\rm m/s}$.
Thus $v'\approx 0.001-0.01 \ {\rm m/s}$ for styrene-butadiene compounds sliding on different surfaces (which probably are contaminated by
molecules from the rubber compound itself)\cite{exp1}, and $v'\approx 0.01-0.1 \ {\rm m/s}$ has been observed
for PDMS sliding on passivated glass surfaces\cite{Chaud}. The slightly larger velocity $v'$ for the PDMS-glass system
could result from the inert nature of the PDMS molecules and the passivated
glass surfaces used in the experiments; this result in a high PDMS chain mobility,
and to low energy barriers for chain rearrangements, and to short
relaxation time $t'$.

We believe that $t'$-relaxation processes, associated with chain rearrangements in the substrate potential field,
may be the origin of the velocity dependency of the contact area observed by Vorvolakos and Chaudhury\cite{Chaud}
when sliding PDMS spheres on glass surfaces.
However, one counter argument is that the transition of a (nanometer-sized) rubber patch from the bound commensurate-like state, to an
incommensurate-like configurations, which can easily slip before the next attachment,
may involve only a small increase in the local surface separation.
Since the surface separation is only slightly increased, the rubber-substrate binding energy is only
slightly reduced (but the lateral barrier for sliding strongly reduced) in the incommensurate state\cite{smooth}.
However, in the detached state thermal fluctuations may result in an entropic repulsion
like observed in Fig. \ref{T0T20}, which could result in a larger average surface separation then otherwise expected,
and to a negligible adhesion energy in the detached state. In fact, the elastic energy stored in the elongated contact region just before
detachment may be converted into local heating of the rubber which may favor the detached entropic repulsive state. As the
local temperature decreases due to heat diffusion, the repulsive entropic effect disappears and the rubber patch return to a pinned,
commensurate-like state. This could explain the decrease in the contact area with increasing sliding speed
observed by Vorvolakos and Chaudhury.

\begin{figure}
\centering
\includegraphics[width=0.5\textwidth]{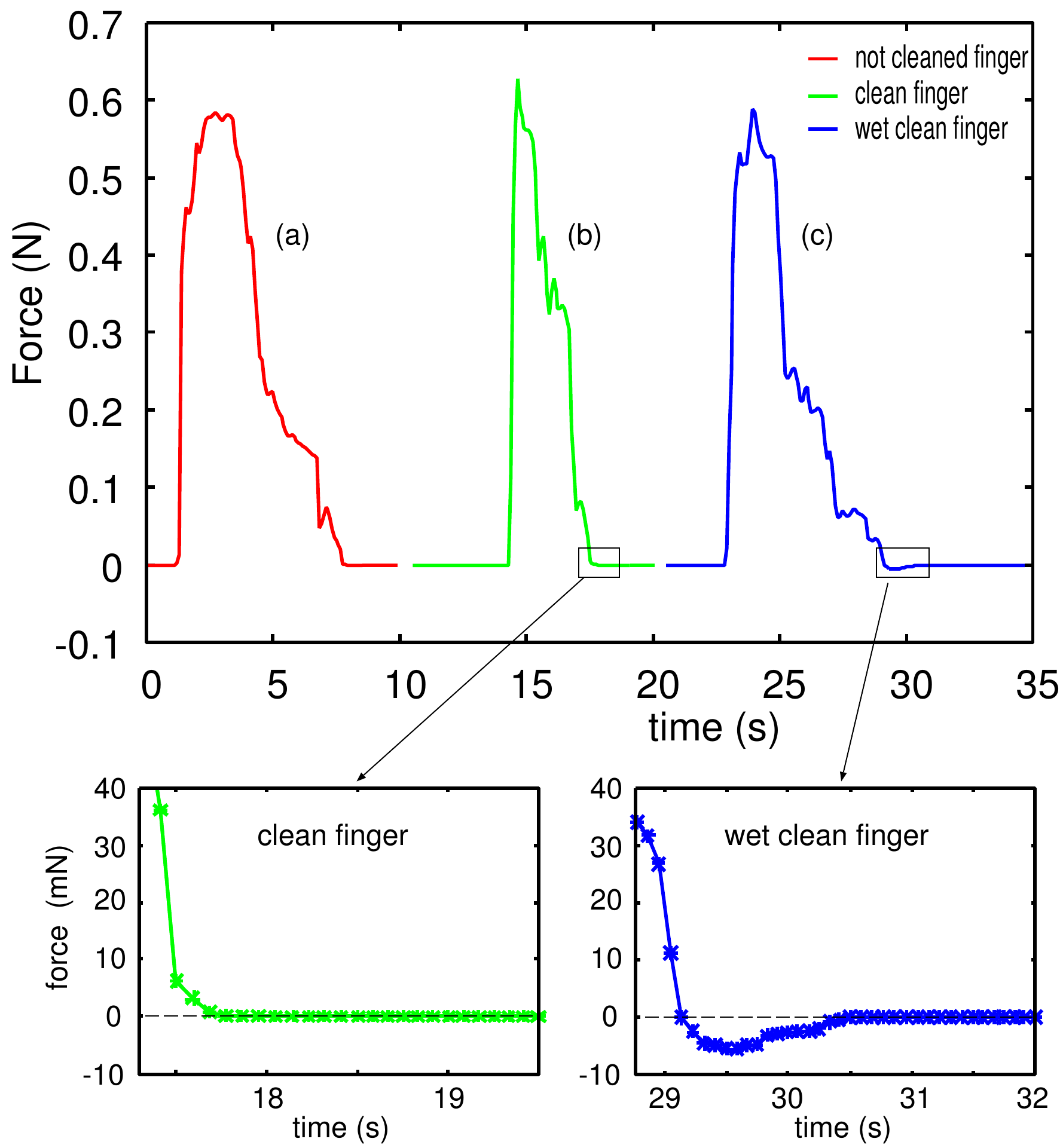}
\caption{\label{fingeradhesion}
The interaction force between a human finger and a dry glass plate cleaned by acetone and isopropanol.
Case (a) (red curve) is for a not cleaned finger,
(b) (green curve) for a finger cleaned with soap water and (c) (blue curve) for a clean wet finger.
In case (a) and (b) no (macroscopic) adhesion is observed, while in case (c) do we observe
adhesion with a pull-off force $F_{\rm pull-off} \approx 5.5 \ {\rm mN}$.}
\end{figure}

\vskip 0.3cm
{\bf 4. Finger-glass adhesion experiments}

Several recent experimental studies have shown that when a tangential force is applied to a  human finger squeezed against a
flat glass surface, the glass-finger nominal contact area decreases\cite{PNAS,Finger1}. This has been tentatively explained using the
adhesion theory described in Sec. 2.1 (see Ref. \cite{Cia}). However, we have performed adhesion experiments
for a finger in contact with a glass plate, and for a dry finger we do not observe any macroscopic adhesion so the explanation
proposed in Ref. \cite{Cia} cannot explain the observed
decrease in the contact area with increasing tangential force.

Fig. \ref{fingeradhesion}
shows the interaction force between a human finger and a dry glass plate cleaned by acetone and isopropanol.
Case (a) (red curve) is for a not cleaned finger,
(b) (green curve) for a finger cleaned with soap water and (c) (blue curve) for a clean wet finger.
In case (a) and (b) no (macroscopic) adhesion is observed, while in case (c) do we observe
adhesion with a pull-off force $F_{\rm pull-off} \approx 5.5 \ {\rm mN}$.
This is similar to what is expected if a capillary bridge is formed between the glass surface and the finger.
Thus for a thick water film $F_{\rm pull-off} \approx  4 \pi R \gamma$,
where the water surface tension $\gamma \approx 0.07 \ {\rm J/m^2}$ and $R$ is the radius of curvature of the finger.
If we use $R\approx 0.7 \ {\rm cm}$ we obtain the observed pull-off force.
However, the pull-off force depends on the volume of water on the finger and if the water volume is too small (less then $\sim 1 \ {\rm mm}^3$)
no adhesion is observed which we interpret as resulting from the skin surface roughness and the elastic rebound of the deformed skin.

In our MD simulations without adhesion for the contact of an elastic slab with a rigid corrugated substrate we observed a small
increase in the contact area with increasing sliding speed corresponding to increasing lateral force. Thus the contact area increased
by about $10\%$ as the friction coefficient increased from $\approx 0.01$ to $\approx 0.1$.

We believe that the reduction in the contact area observed for the human finger with increasing lateral force
is due to the complex inhomogeneous (layered) nature of the finger and to the large deformations involved.
It is also possible that the superposition of the normal and parallel deformation fields assumed in most analytic treatments
is not accurate enough when the parallel deformations becomes large and coupling effects becomes important.
This conclusion is supported by finite element calculations performed by Mergel et al\cite{Merg1} (see also \cite{Merg2}
and \cite{replica}), 
which shows that even without adhesion there is a the reduction in the contact area between an elastic cylinder 
and a flat surface as a tangential force is applied to the cylinder.

\vskip 0.3cm
{\bf 5 Summary and conclusion}

We have used molecular dynamics to study the contact between an elastic slab with a flat surface and a rigid, cylinder corrugated substrate.
We have considered cases with and without adhesion. The most important results are:

(1) For low sliding speeds the contact width is found to be nearly velocity independent, while for high speeds it decreases when
adhesion is included, and increases without adhesion.

(2) When adhesion is included the contact is asymmetric, extending further on the opening crack side then on the closing crack side.
We attribute this to lattice pinning: on the opening crack side the crack tip perform stick-slip motion, where
atoms snap out of contact in rapid events, followed by time periods where the crack tip is pinned. In the rapid
slip events elastic waves (phonons) are emitted from the crack tip resulting in a larger crack propagation energy then the
adiabatic value.

(3) Adhesion experiments between a human finger and a flat smooth glass surface were carried out. We found
that there was no macroscopic adhesion between these contacting pairs in the dry state.
Based on this result, we suggest that the decrease in the contact area
as reported in literature \cite {PNAS,Finger1, PRL} results from non-adhesive contact mechanics,
involving large deformations of a complex layered material.

\vskip 0.3cm
{\bf Acknowledgments}
B. Persson thank G. Carbone for extensive discussions related to the 
contact mechanics calculation presented by Menga, Carbone and Dini  
in Ref. \cite{Menga}. J. Wang would like to thank scholarship from China Scholarship Council (CSC) and funding by National Natural Science Foundation of China (NSFC): grant number U1604131.

\vskip 0.3cm
{\bf Appendix A: Contact stiffness}

When a tangential stress act in a circular contact region (area $A_0=\pi r_0^2$) on the surface of an elastic half-space, the contact region will
displace relative to the solid far away from the contact center by ${\bf u}({\bf x})$. Here ${\bf x}=(x,y)$ is the coordinate of a material
point on the (undeformed) surface of the elastic half-space.
We assume that the stress act along the $x$-direction and define the average displacement
$$ u = {1\over A_0} \int_{A_0} d^2x \ u_x({\bf x})\eqno(A1)$$
Using the theory of elasticity
$$u_x({\bf x}) = {1\over 2 \pi G} \int_{A_0} d^2x' \ \tau({\bf x'})
\left ({1-\nu \over |{\bf x}-{\bf x}'|}+{\nu (x-x')^2\over  |{\bf x}-{\bf x}'|^3}\right ) \eqno(A2)$$
If we assume that $\tau({\bf x})=\tau$ is constant
one can show from (A1) and (A2) that\cite{Am,Menga,Kim}
$$k u = F_x$$
where $F_x = \tau A_0$, and where $k= \alpha r E^*$ where\cite{Am} 
$\alpha = (1-\nu)/(A+B\nu)$ with (from numerical integration) $A\approx 0.54$ and $B\approx -0.27$.
For $\nu = 0.5$ Menga et al\cite{Menga} showed that $\alpha = \pi^2/8$, and 
McMeeking et al\cite{Kim} later obtained for arbitrary $\nu$ the analytic
result $\alpha = (3\pi^2 /8)(1-\nu)/(2-\nu)$, which is consistent with the numerical values 
for $A$ and $B$ given above.

In a similar way one can calculate\cite{Menga}
$$\bar u_x = {1\over 2 \pi} \int_0^{2 \pi} d\phi \ u_x(r {\rm cos}\phi, r {\rm sin}\phi )$$
and show that $\beta r E^* \bar u_x = F_x$ with $\beta = 4\alpha/3$.

If instead of a constant shear stress one assume a constant displacement (which require a shear stress proportional to
$[1-(r/r_0)^2]^{-1/2}$) one obtain\cite{Jons} $k'  u = F_x$ with $k'=\alpha' r E^*$ with $\alpha' = 4(1-\nu)/(2-\nu)
=(1-\nu)/(A'+B'\nu)$ with $A'=0.5$ and $B'=-0.25$, i.e., very close to the result when the shear stress is constant.

\end{document}